\documentclass[aps,physrev,twocolumn,groupedaddress]{revtex4-2}

\usepackage{siunitx}
\usepackage{tikz}
\usetikzlibrary{calc}
\usepackage{booktabs}
\usepackage{float}
\usepackage{amsmath}
\usetikzlibrary{shapes.geometric, arrows, positioning,shapes.multipart}
\usepackage{amssymb}
\usepackage{nicefrac}

\renewcommand{\epsilon}{\varepsilon}

\newcommand{\nutensor}{\ensuremath{\overline{\overline{\nu}}}}
\newcommand{\sigmatensor}{\ensuremath{\overline{\overline{\sigma}}}}

\renewcommand{\d}{\ensuremath{ \mathrm{d}} }

\newcommand{\A}{\ensuremath{\vec{A}}}
\newcommand{\AvecComplex}{\ensuremath{\underline{\vec{A}}}}

\newcommand{\Js}{\ensuremath{\vec{J}_{\mathrm{s}}}}

\newcommand{\JsvecComplex}{\ensuremath{\underline{\vec{J}}_{\mathrm{s}}}}

\newcommand{\nuComp}{\ensuremath{\underline{\nu}}}

\newcommand{\adjustedaccent}[1]{%
	\mathchoice{}{}
	{\mbox{\raisebox{-.5ex}[0pt][0pt]{$\scriptstyle#1$}}}
	{\mbox{\raisebox{-.35ex}[0pt][0pt]{$\scriptscriptstyle#1$}}}
}

\newcommand{\adjustedaccentsecondbow}[1]{%
	\mathchoice{}{}
	{\mbox{\raisebox{-.7ex}[0pt][0pt]{$\scriptstyle#1$}}}
	{\mbox{\raisebox{-.35ex}[0pt][0pt]{$\scriptscriptstyle#1$}}}
}

\newcommand\bow[1]{\overset{\adjustedaccent{\smallfrown}}{#1}}
\newcommand\secondbow[1]{\overset{\adjustedaccentsecondbow{\smallfrown}}{#1}}

\newcommand{\ahatc}[1]{\ensuremath{\bow{\underline{\mathbf{a}}}_{#1}}} 

\newcommand{\ff}{\ensuremath{f_{\mathrm{f}}}}
\newcommand{\wf}{\ensuremath{\omega_{\mathrm{f}}}}

\definecolor{verylightblue}{RGB}{0,156,218}
\definecolor{verylightpink}{RGB}{201,48,142}
\definecolor{red}{RGB}{230,0,26}
\definecolor{green}{RGB}{153,192,0}
\definecolor{blue}{RGB}{0,90,169}
\definecolor{lightblue}{RGB}{0,131,204}
\definecolor{lightsmaragd}{RGB}{0,131,204}
\definecolor{orange}{RGB}{236,101,0}
\definecolor{lightpurple}{RGB}{166,0,132}

\definecolor{darkblue_one}{RGB}{0,78,138} 
\definecolor{darkblue_two}{RGB}{0,104,157} 

\definecolor{darkemerald}{RGB}{0,136,119} 
\definecolor{darkgreen_one}{RGB}{127,171,22} 
\definecolor{darkgreen}{RGB}{127,171,22} 

\definecolor{darkgreen_two}{RGB}{177,189,0} 
\definecolor{darkyellow_one}{RGB}{215,172,0} 
\definecolor{darkyellow_two}{RGB}{210,135,0} 
\definecolor{darkyellow}{RGB}{210,135,0} 

\definecolor{darkorange}{RGB}{204,76,3} 

\definecolor{darkred}{RGB}{185,15,34} 

\definecolor{darkpurple}{RGB}{149,17,105} 

\tikzstyle{start} = [rectangle, rounded corners, minimum width=3.5cm, minimum height=0.3cm, text centered, draw=darkred, line width = 0.6mm]
\tikzstyle{stop} = [rectangle, rounded corners, minimum width=3.5cm, minimum height=0.5cm, text centered, draw=darkblue_two, line width = 0.6mm]
\tikzstyle{process_blue} = [rectangle, minimum width=3.5cm, minimum height=1cm, text centered, draw=darkblue_two, line width = 0.6mm]
\tikzstyle{process_red} = [rectangle, minimum width=3.5cm, minimum height=1cm, text centered, draw=darkred, line width = 0.6mm]
\tikzstyle{process_black} = [rectangle, minimum width=3.5cm, minimum height=1cm, text centered, draw=black, line width = 0.6mm]
\tikzstyle{decision} = [diamond, aspect=2.5, minimum width=0.8cm, minimum height=0.1cm, text centered, draw=darkblue_two, line width = 0.6mm]
\tikzstyle{arrow} = [thick,->,>=stealth]
\begin{document}


\title{Experimental Validation of HomHBFEM Simulations of Fast Corrector Magnets for PETRA IV}


\author{Jan-Magnus Christmann, Laura Anna Maria D'Angelo, and Herbert De Gersem}
\affiliation{Institute for Accelerator Science and Electromagnetic Fields, Technical University of Darmstadt, Darmstadt, Germany}
\author{Sven Pfeiffer, Sajjad Hussain Mirza, Adeel Amjad, Lucas Rousselange, and Matthias Thede}
\affiliation{Deutsches Elektronen-Synchrotron DESY, Hamburg, Germany}


\begin{abstract}
This paper presents experimental validation of the homogenized harmonic balance finite element method (HomHBFEM), which we have developed as a dedicated simulation technique for magnets with fast excitation cycles, in particular the fast corrector (FC) magnets for PETRA IV at DESY. The HomHBFEM allows efficient three-dimensional nonlinear eddy-current simulations of laminated magnets at elevated frequencies with a relatively coarse finite element (FE) mesh and without computationally expensive time-stepping. This is achieved by combining a frequency-domain-based homogenization technique with the harmonic balance FE method. The simulation results for the magnetic flux density along the axis of the FC magnets as a function of frequency and the resulting integrated transfer function (ITF) are compared to Hall probe and search coil measurements of the first prototype FC magnet for PETRA IV. A good agreement between simulated and measured ITFs is achieved for excitation frequencies from $\SI{10}{\hertz}$ to $\SI{10}{\kilo\hertz}$.

\end{abstract}


\maketitle

\section{Introduction}\label{sec:intro}
The upgrade of \mbox{PETRA III} at DESY to the new fourth generation synchrotron radiation (SR) source \mbox{PETRA IV} is currently being planned. PETRA IV aims to become the world's brightest storage ring-based SR source. Therefore, an electron beam of ultra-low emittance must be provided~\cite{schroer_2019aa, schroer_2022aa, bartolini_2022aa}. The goal is to achieve a horizontal and vertical emittance of $\SI{12}{\pico\meter \radian}$. Given these ultra-low emittances and the correspondingly small beam size, meeting the typical beam stability requirement of at least $\SI{10}{\percent}$ of the beam size~\cite{kallakuri_2019aa, broucquart_2022aa, mirza_2023aa} presents a significant challenge. Hence, a fast orbit feedback (FOFB) system is needed to counteract disturbances up to the kilohertz range. 

\begin{figure}[t]
  \centering
  \includegraphics[width=\columnwidth]{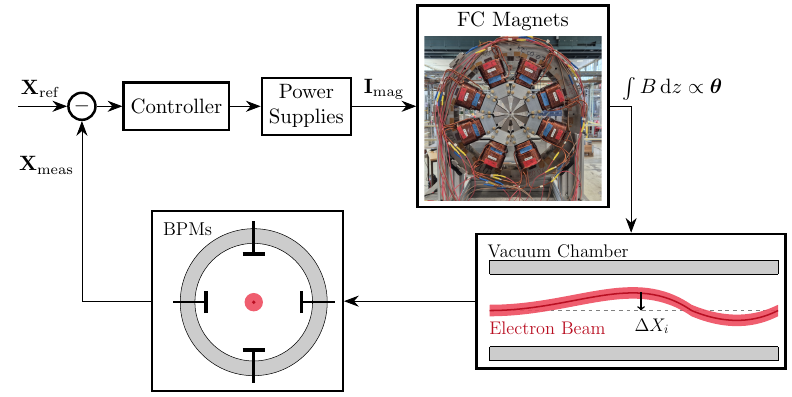}
  \caption{Schematic control loop of the FOFB system.}
  \label{fig:fofb_loop}
\end{figure}

A control loop of a FOFB system~\cite{tavares_2023aa, zhu_2023aa,kallakuri_2024aa} is sketched in Fig.~\ref{fig:fofb_loop}. The transversal positions of the electron beam along the ring are measured via beam position monitors (BPMs) and summarized in the vector $\mathbf{X}_{\mathrm{meas}}$. The difference between $\mathbf{X}_{\mathrm{meas}}$ and the reference values $\mathbf{X}_{\mathrm{ref}}$, i.e., the beam's deviation from the design orbit, is fed back to a controller. The controller determines the setpoints of the power supplies, exciting the fast corrector (FC) magnets via the supplied currents summarized in the vector $\mathbf{I}_{\mathrm{mag}}$. Each of the FC magnets then provides an integrated magnetic flux density, which, after additional attenuation by the vacuum chamber, deflects the beam by angles $\boldsymbol{\theta}$. These deflection angles are chosen by a correction algorithm to minimize the deviation from the design orbit along the ring~\cite{chao_2013aa}.

For the design of the controller, we need to know the dynamic behavior of all components in Fig~\ref{fig:fofb_loop}. In this study, we focus on the FC magnets. Specifically, we are interested in the relationship between the currents $\mathbf{I}_{\mathrm{mag}}$ and the integrated magnetic flux densities provided by the FC magnets as a function of frequency. Since the FOFB for PETRA IV needs to reach a disturbance rejection bandwidth of $\SI{1}{\kilo\hertz}$~\cite{mirza_2023aa}, we must analyze the behavior of the FC magnets for frequencies up to the kilohertz range. At such elevated frequencies, eddy-current effects in the magnet yokes play a major role despite their laminated structure.
To investigate the eddy-current effects, we carried out extensive finite-element (FE) simulation studies  before the first prototype was built~\cite{christmann_2024aa, christmann_2024ab}. Since the magnets are short compared to their transversal dimensions, a three-dimensional (3D) analysis was indispensable. Such 3D simulations of laminated magnets are challenging because the FE mesh not only has to resolve the individual laminations, which usually have a thickness of $\SI{1}{\milli\meter}$ or less, but also the skin depth~\cite{meunier_2008aa}, which scales with the frequency $f$ according to $\nicefrac{1}{\sqrt{f}}$. In a brute-force approach, without the use of special techniques, this leads to an enormous amount of degrees of freedom (d.o.f.) and thus a prohibitive computational effort~\cite{sabariego_2020aa}. 

We developed a dedicated simulation strategy based on the frequency-domain-based homogenization technique proposed in~\cite{dular_2003aa}. The homogenization technique replaces the laminations with a bulk model characterized by frequency-dependent material tensors, thus allowing us to use a much coarser FE mesh. In its original form, however, the homogenization technique is only applicable under the assumption of linear magnetic material properties, i.e., a constant permeability has to be specified for the yoke. To lift this limitation, we combined the homogenization technique with a harmonic balance finite element method (HBFEM)~\cite{yamada_1988aa, yamada_1989aa, yamada_1991aa}, which introduces a multiharmonic approach, i.e., the solution is represented as a truncated Fourier series. With the resulting homogenized harmonic balance finite element method (HomHBFEM), we can simulate the FC magnets with a relatively coarse mesh, and include a nonlinear $B$--$H$ curve without costly time stepping. The HomHBFEM was recently published by us together with extensive numerical verification studies, but without experimental validation~\cite{christmann_2025aa}. 

In this paper, we compare simulations and measurements of the first prototype magnet for PETRA IV. The good agreement between simulation and measurement provides the first experimental validation of the HomHBFEM. The paper is structured as follows. First, in Sec.~\ref{sec:fc_magnets}, we give more detailed information about the FC magnets for PETRA IV. Then, we briefly present the measurement setups in Sec.~\ref{sec:meas_setup} and give an explanation of the simulation method in Sec.~\ref{sec:sim_method}. Finally, the main contribution of this paper, the validation of the HomHBFEM with the measurements, is presented in Sec.~\ref{sec:meas_vs_sim}. Finally, Sec.~\ref{sec:conclusion} concludes the paper.

\section{Fast Corrector Magnets for PETRA IV}\label{sec:fc_magnets}
\begin{figure}
     \includegraphics[width=0.45\linewidth]{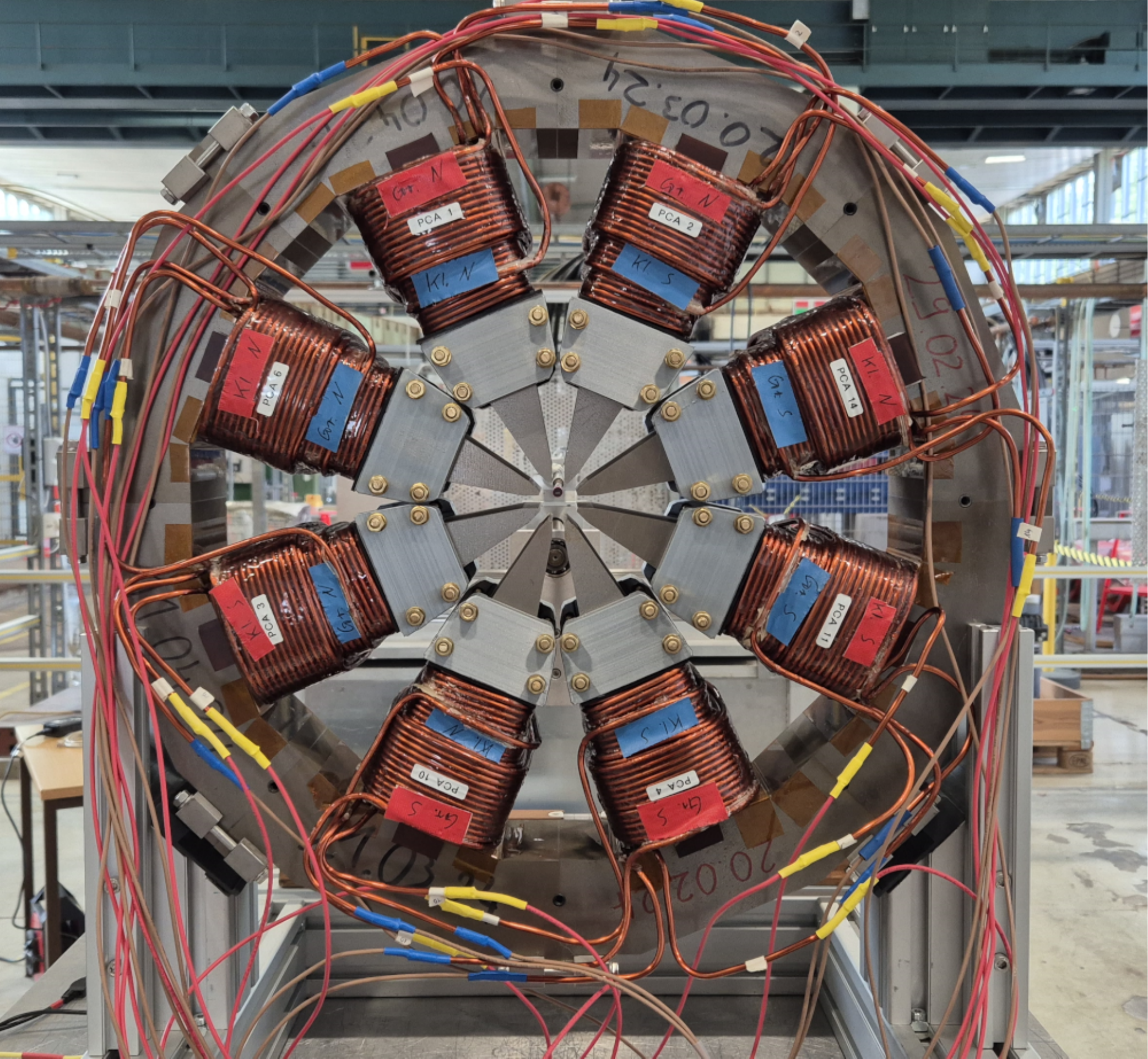}
    \includegraphics[width=0.43\linewidth]{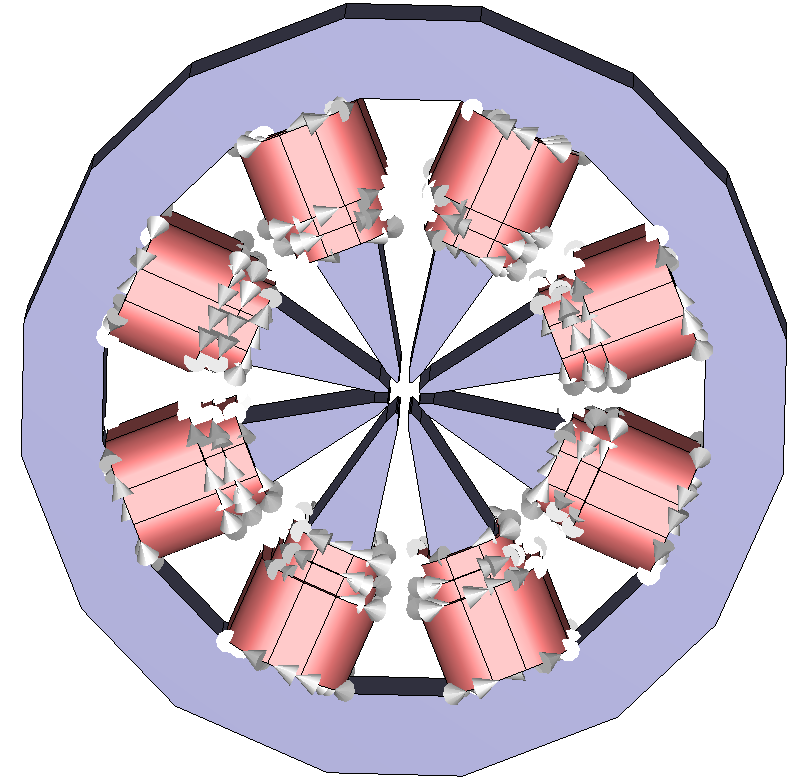} 
    \caption{Left: Prototype FC magnet. Right: Simulation model.}
    \label{fig:fc_magnet_1}
\end{figure}

\begin{figure}
     \includegraphics[width=0.3\linewidth]{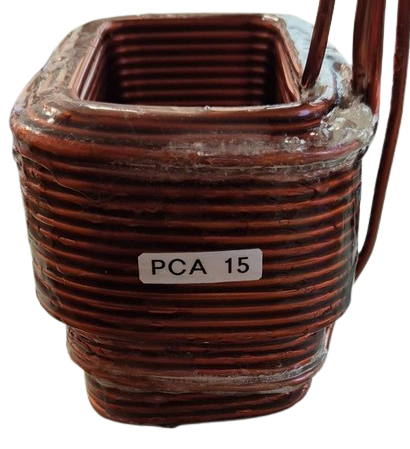}\quad
    \includegraphics[width=0.25\linewidth]{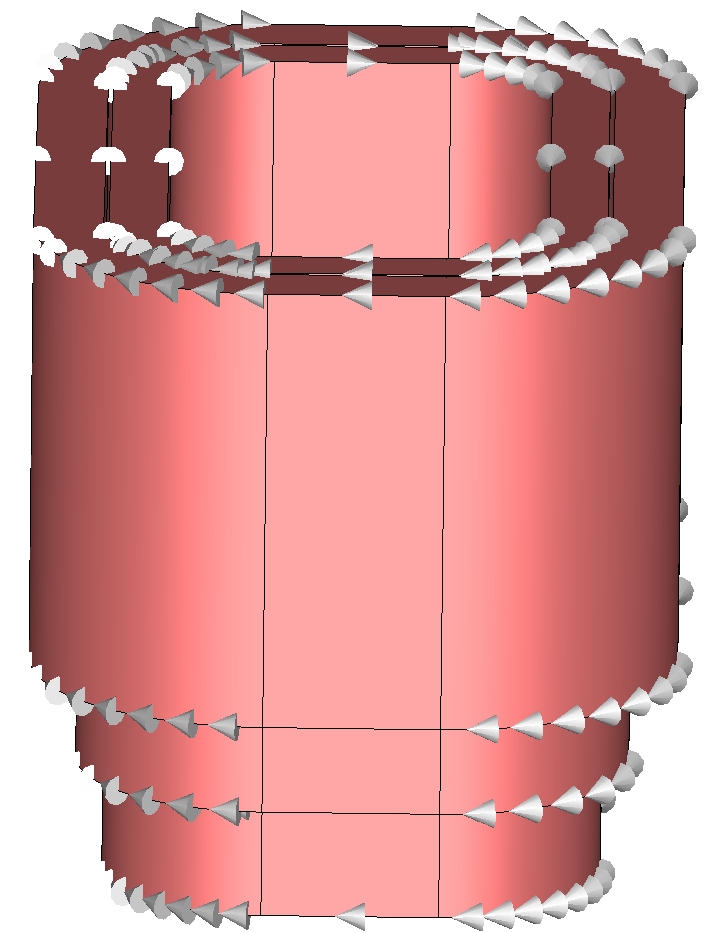} 
    \caption{Left: Coil of prototype FC magnet. Right: Simulation model.}
    \label{fig:fc_magnet_3}
\end{figure}

The FC magnets for PETRA IV use an 8-pole design, very similar to the one used by the FC magnets for the Advanced Photon Source Upgrade at ANL~\cite{depaola_2018aa}. A picture of the first prototype alongside the model used for the simulations is shown in Fig.~\ref{fig:fc_magnet_1}, the geometric parameters are summarized in Table~\ref{tab:corrector_geo_params}. The FC magnets will be used for DC corrections up to $\theta = \SI{560}{\micro\radian}$ and smaller AC corrections at frequencies up to several kilohertz, e.g., a maximum of $\theta = \SI{30}{\micro\radian}$ below $\SI{100}{\hertz}$ and $\theta = \SI{3}{\micro\radian}$ at $\SI{1}{\kilo\hertz}$.
\begin{table}
	\caption{Geometric parameters of the prototype FC magnet.}\label{tab:corrector_geo_params}
	\centering
	\begin{tabular}{ll}
		\toprule
		\toprule
		Yoke length & $\SI{86}{\milli\meter}$   \\
		Yoke diameter &$\SI{564}{\milli\meter}$   \\
		Aperture diameter  & $\SI{25}{\milli\meter}$ \\
		Insertion length & $\SI{141}{\milli\meter}$ \\
		\bottomrule
		\bottomrule
	\end{tabular}
\end{table}
The 8-pole design allows to combine vertical and horizontal corrector in one relatively compact magnet with a good field quality. Each pole is equipped with a main and auxiliary coil, whose turn ratio is chosen such that the sextupole and decapole multipole components of the aperture field are canceled entirely. To that end, a layered coil structure is used, see Fig.~\ref{fig:fc_magnet_3}. 
The two outer layers make up the auxiliary coil with a total of $27$ turns and the inner layer constitutes the main coil with $65$ turns.  The maximum DC current for both auxiliary and main coils is $\SI{15}{\ampere}$, which leads to an aperture dipole field of roughly $\SI{90}{\milli\tesla}$.
Furthermore, the design features the possibility of including a skew quadrupole in the same magnet by adding additional coils. This option could be explored in the future, but at the moment no skew quadrupole is included, i.e., only a dipole field is provided, either in vertical or in horizontal direction, depending on which set of coils is powered. 

\begin{figure}
  \centering
  \includegraphics[width=0.48\linewidth]{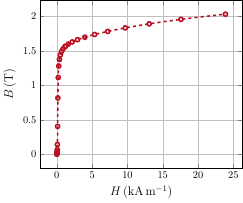}\hfill
  \includegraphics[width=0.48\linewidth]{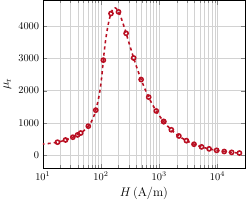}
  \caption{Left: $B$--$H$ curve powercore$^{\text{\textregistered}}$ 1400-100AP. Right: Relative permeability of powercore$^{\text{\textregistered}}$ 1400-100AP.}
  \label{fig:fc_magnet_2}
\end{figure}

The yoke material of the prototype magnet is powercore$^{\text{\textregistered}}$ 1400-100AP, a non-grain-oriented electrical steel grade manufactured by thyssenkrupp~\cite{powercore}, with a lamination thickness of $d = \SI{1}{\milli\meter}$. The stacking factor $\gamma$, i.e., the percentage of the yoke's volume consisting of steel is roughly $\SI{98}{\percent}-\SI{99}{\percent}$. The conductivity of the laminates was specified by the manufacturer as \mbox{$\sigma = \SI{5.814}{\mega\siemens\per\meter}$}. The $B$--$H$ curve was measured at DESY with a split-coil permeameter and is shown in Fig.~\ref{fig:fc_magnet_2} together with the corresponding relative permeability as a function of the magnetic field \mbox{strength $H$}. The other material parameters are summarized in Table~\ref{tab:corrector_mat_params}.

\begin{table}
	\caption{Material parameters of the prototype FC magnet yoke.}\label{tab:corrector_mat_params}
	\centering
	\begin{tabular}{ll}
		\toprule
		\toprule
		  Conductivity & $\SI{5.814}{\mega\siemens\per\meter}$  \\
		Stacking factor & $\SI{98}{\percent}-\SI{99}{\percent}$  \\
		  Lamination thickness  & $\SI{1}{\milli\meter}$ \\
		\bottomrule
		\bottomrule
	\end{tabular}
\end{table}

\section{Measurement Setups}\label{sec:meas_setup}
\subsection{Hall Sensor Measurement}
\begin{figure}
    \centering
     \includegraphics[width=0.45\linewidth]{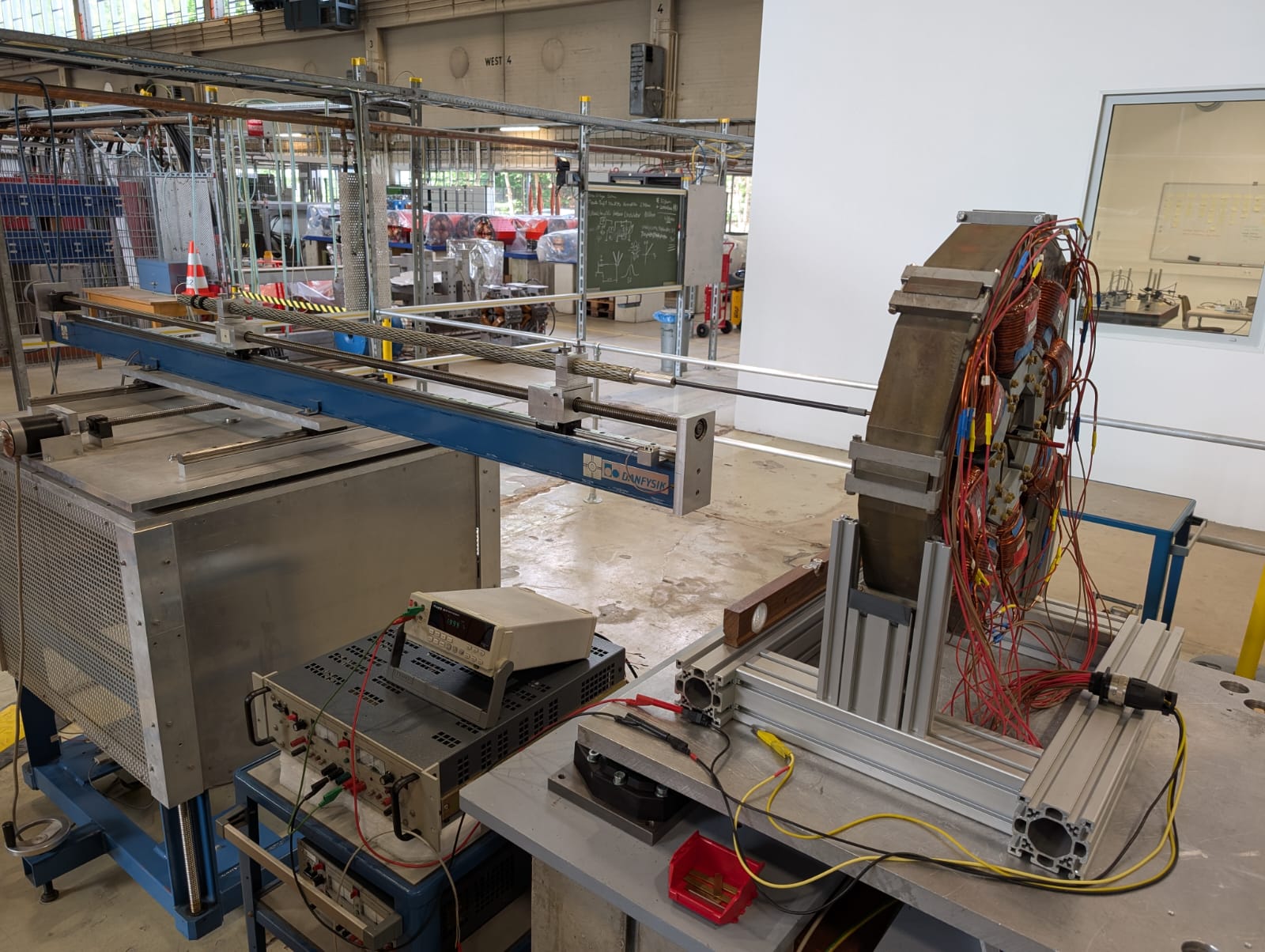}\quad
    \includegraphics[width=0.45\linewidth]{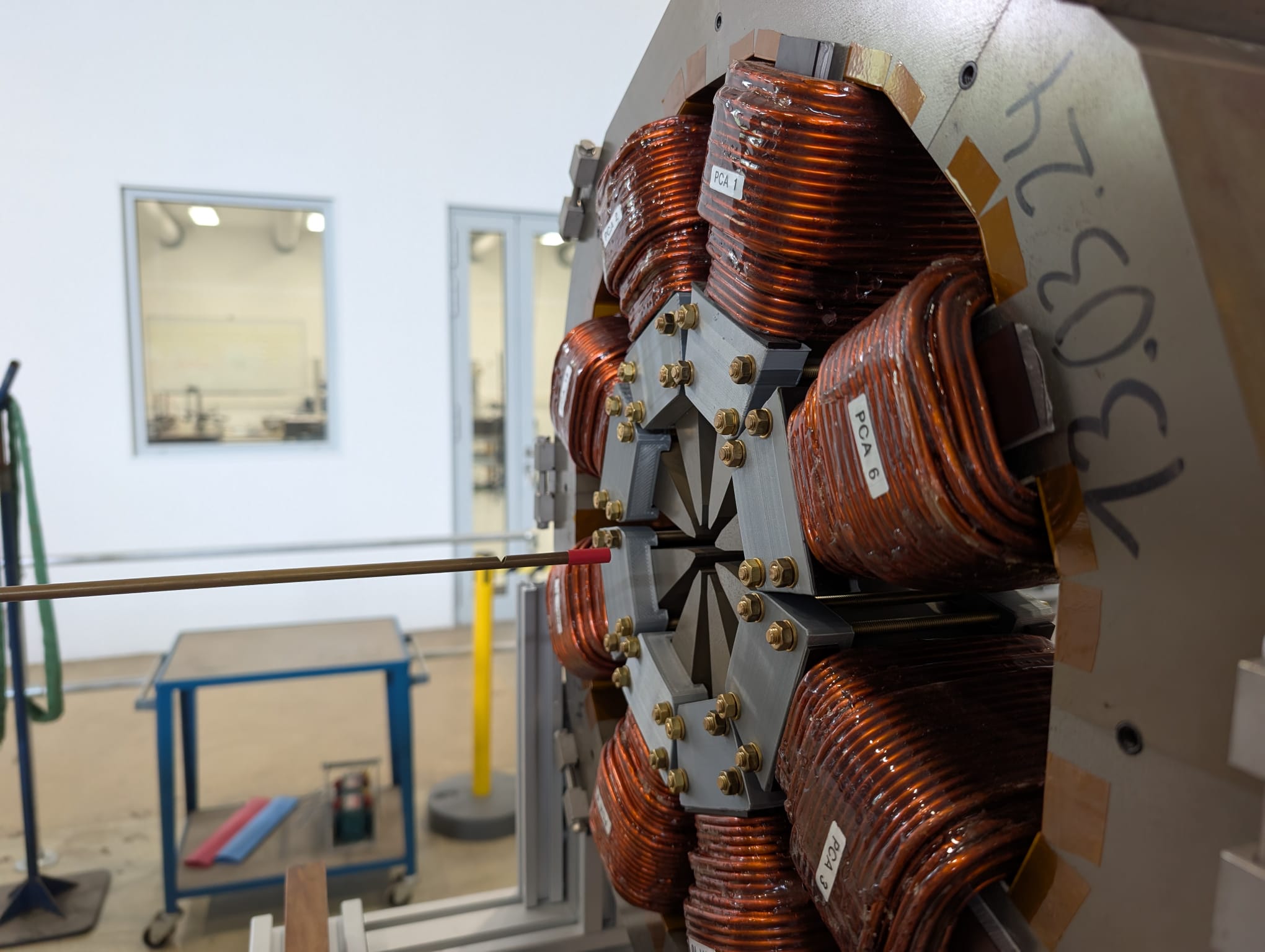} 
    \caption{First measurement setup with a Hall probe on a movable stage for position-dependent measurement of the magnetic flux density.}
    \label{fig:measurement_setup_1}
\end{figure}

In the first measurement setup, shown in Fig.~\ref{fig:measurement_setup_1}, we measured the magnetic flux density along the magnet axis with a Hall probe. The Hall probe is positioned on a movable stage and connected to a magnetic field transducer (SENIS F3A~\cite{senis_transducer}) with a $\SI{3}{\decibel}$ bandwidth of $\SI{25}{\kilo\hertz}$.

The Hall probe was moved along the axis in steps of $\SI{2}{\milli\meter}$ from $z = \SI{-150}{\milli\meter}$ to $z=\SI{150}{\milli\meter}$. At each position, the magnetic flux density was recorded in magnitude and phase as a function of frequency.

\begin{figure}
    \centering
    \includegraphics[width=0.65\linewidth]{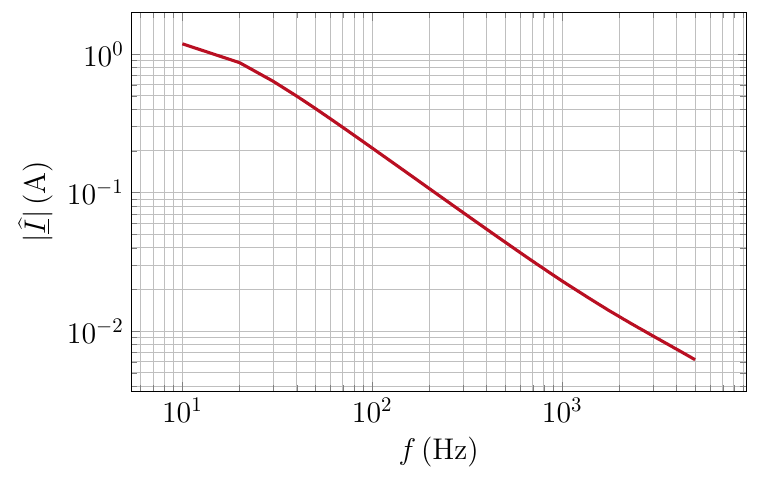}
    \caption{Coil current of FC magnet as a function of frequency during Hall probe measurement.}
    \label{fig:hall_sensor_meas_current}
\end{figure}

In this measurement setup, the waveform generator (AD Pro 3450 by Digilent~\cite{ad_pro}) is connected to a power amplifier (BOP 20-20M by KEPCO) such that we can drive higher AC currents in the FC magnet coils. The magnitude of the current in the FC magnet coils as a function of frequency during this measurement is shown in Fig.~\ref{fig:hall_sensor_meas_current}.
\subsection{Search Coil Measurement}\label{sec:measurement_setup_search_coil}
\begin{figure}
    \centering
    \includegraphics[width=0.45\linewidth]{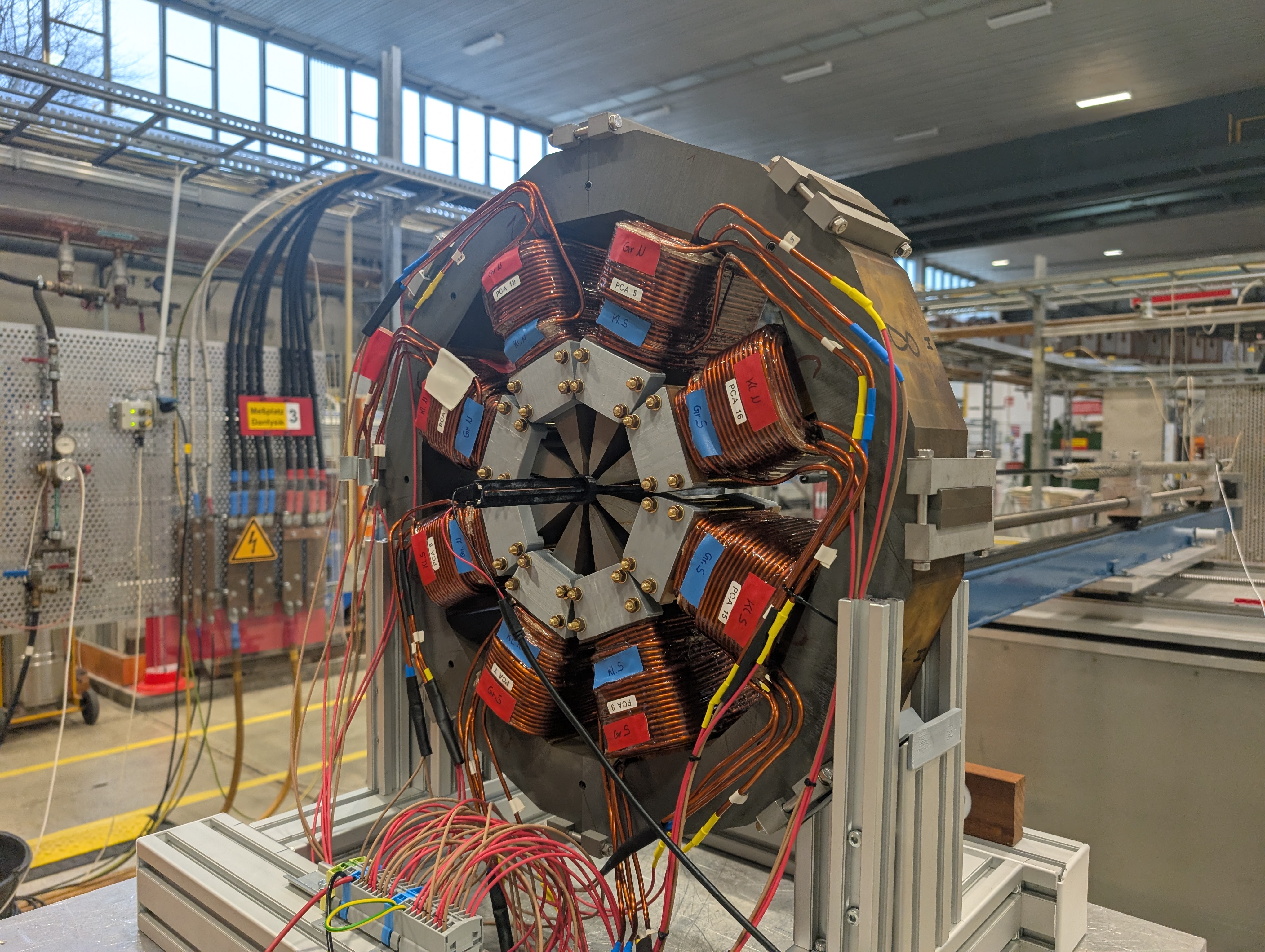}
    \includegraphics[width=0.45\linewidth]{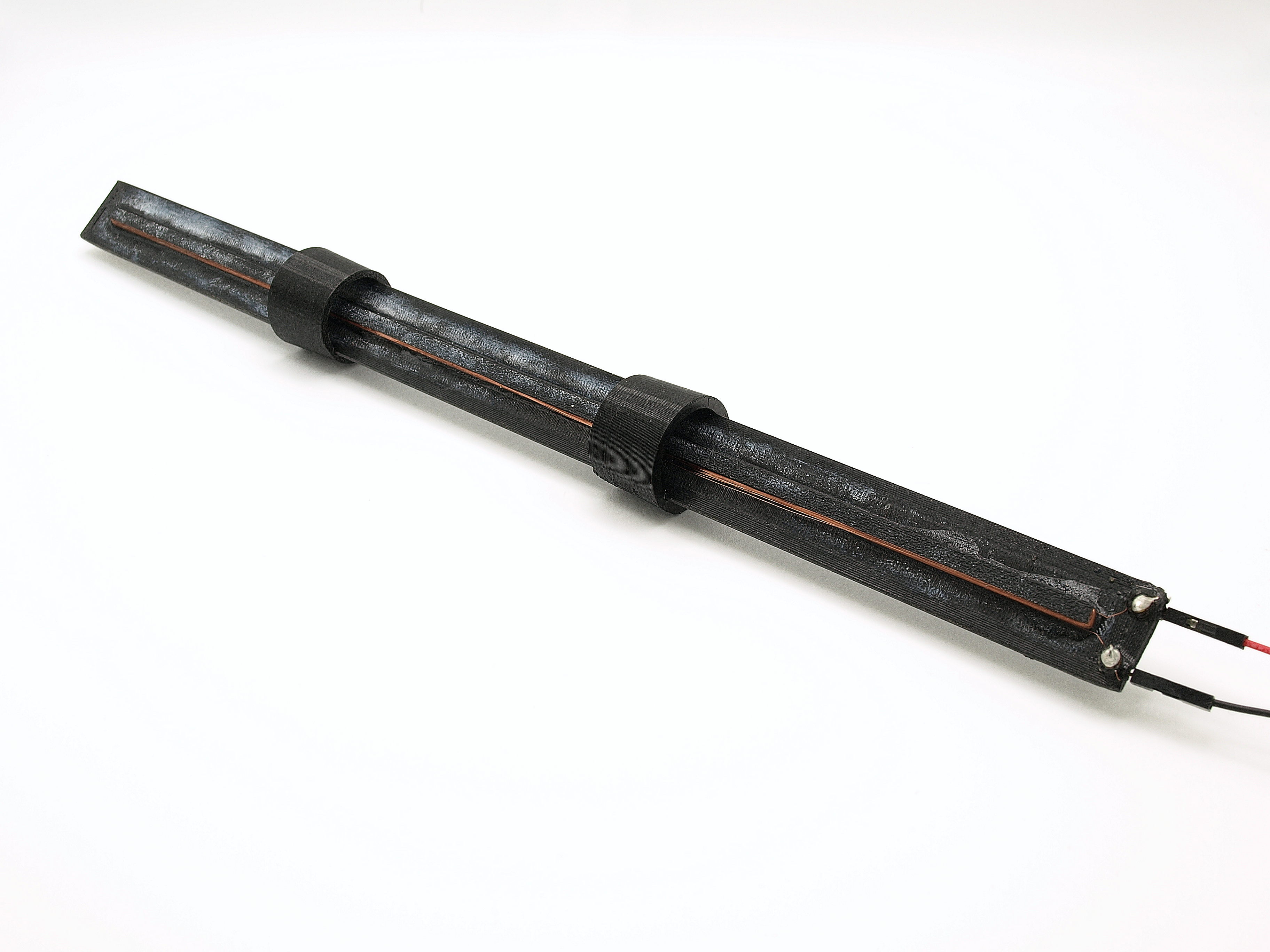} 
    \caption{Second measurement setup with search coil.}
    \label{fig:measurement_setup_2}
\end{figure}

In the second measurement setup, shown in Fig.~\ref{fig:measurement_setup_2}, we measure the induced voltage in a search coil to compute the integrated magnetic flux density along the magnet axis as a function of frequency. To that end, the search coil is inserted in the center of the aperture such that the magnetic flux density vector $\vec{\underline{B}}$ produced by the FC magnet is perpendicular to the search coil cross section. 

The integrated magnetic flux density $\underline{B}_{\mathrm{int}}$ is computed from the induced voltage $\underline{U}_{\mathrm{ind}}$ as follows. According to Faraday-Lenz's law, the induced voltage in a  search coil with cross-sectional area $A_{\mathrm{coil}}$ is
\begin{align}\label{eq:lenz_law}
    \notag \underline{U}_{\mathrm{ind}} &= -j \omega N_{\mathrm{t}} \underline{\Phi} = -j \omega N_{\mathrm{t}}  \int_{A_{\mathrm{coil}}} \underline{\vec{B}} \cdot \mathrm{d}\vec{A} \\  &= -j \omega N_{\mathrm{t}}  w \int_{l_z} \underline{B}\,\mathrm{d}z = - j\omega N_{\mathrm{t}}  w \underline{B}_{\mathrm{int}}, 
\end{align}
where $\omega = 2\pi f$ denotes the angular frequency, $\underline{\Phi}$ the magnetic flux through the search coil, $N_{\mathrm{t}}$ the number of search coil turns and $w$ and $l_z$ are the width and the length of the search coil. Hence, $\underline{B}_{\mathrm{int}}$ can be determined via 
\begin{equation}
    \underline{B}_{\mathrm{int}} = -\frac{\underline{U}_{\mathrm{ind}}}{j \omega N_{\mathrm{t}}  w}.
\end{equation}
Note that Eq.~\eqref{eq:lenz_law} assumes that $\underline{\vec{B}}$ is perpendicular to the coil cross section and does not vary significantly across the coil width~\cite{tumanski_2007aa}. These assumptions are fulfilled in our case.  

Two slightly different search coils were used. Both are in-house printed designs. The first one, shown in Fig.~\ref{fig:measurement_setup_2}, has $N_{\mathrm{t}} =40$ turns, $l_z = \SI{300}{\milli\meter}$ and $w = \SI{3.3}{\milli\meter}$. This search coil will later on be referred to as \mbox{coil A}. The second search coil will be referred to as coil B and has the same length and number of turns as coil A, but its width is slightly smaller, namely $w = \SI{3.0}{\milli\meter}$.

\begin{figure}
    \centering
    \includegraphics[width=0.65\linewidth]{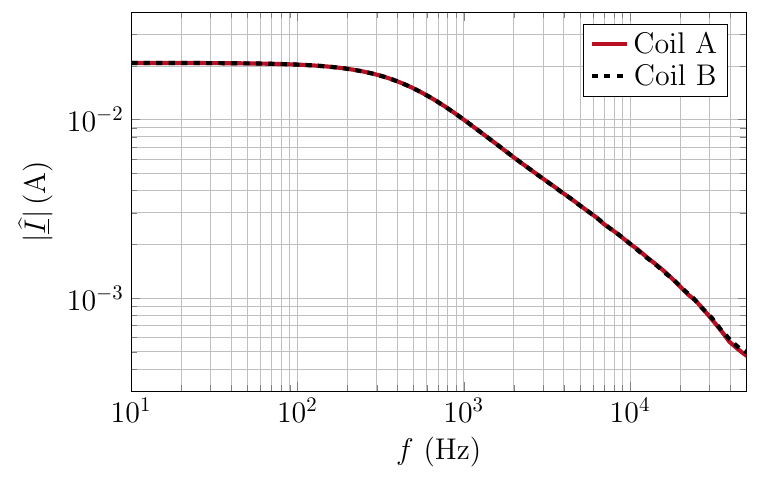}
    \caption{Coil current of FC magnet as a function of frequency during search coil measurements with coil A and coil B.}
    \label{fig:fc_magnet_search_coil_current}
\end{figure}

In this second measurement setup, the waveform generator was directly connected to the FC magnet, without the amplifier from the first measurement setup. As a result, the magnitude of the current in the FC magnet coils is much smaller in this case, see Fig.~\ref{fig:fc_magnet_search_coil_current} and compare to Fig.~\ref{fig:hall_sensor_meas_current}.

\section{Simulation Method}\label{sec:sim_method}
This section provides a brief outline of the developed HomHBFEM. First, its two constituent components, the HBFEM and the homogenization technique, are introduced and then their combination is described. For a detailed explanation and numerical verification studies, we refer the reader to~\cite{christmann_2025aa}.
\subsection{Harmonic Balance Finite Element Method}
To sketch the derivation of the HBFEM~\cite{yamada_1988aa, yamada_1989aa, yamada_1991aa}, we start from the magnetoquasistatic partial differential equation in time domain, which reads
\begin{equation}\label{eq:MQS_NL_TD}
	\nabla \times \left(\nu\left( t \right)  \nabla\times \vec{A}\left(t\right) \right) + \sigma \frac{\partial \vec{A} \left( t \right)}{\partial t} = \vec{J}_{\mathrm{s}}\left( t\right),
\end{equation}
where $\A$ is the magnetic vector potential, $\Js$ the source current density, $\sigma$ denotes the conductivity and $\nu$ the reluctivity. The latter is a function of the time $t$ as its value depends on the magnetic flux density \mbox{$\vec{B}\left(t\right)  = \nabla \times \vec{A}\left( t \right)$}. For brevity, we omit the spatial dependencies.
To carry Eq.~\eqref{eq:MQS_NL_TD} over into frequency domain, we apply the Fourier transform given by 
\begin{equation}
	\underline{\vec{A}}(\omega) = \int_{-\infty}^{\infty} \vec{A}(t) \mathrm{e}^{-\jmath \omega t} \d t
\end{equation}
for the magnetic vector potential and likewise for all other time-dependent quantities. With this definition, the transformation of Eq.~\eqref{eq:MQS_NL_TD} into the frequency domain yields
\begin{equation}\label{eq:MQS_NL_FD}
	\frac{1}{2 \pi} \nabla \times  \left( \nuComp\left(\omega\right) * \nabla \times \AvecComplex \left(\omega\right) \right) + \jmath \omega \sigma \AvecComplex\left( \omega\right)  = \JsvecComplex \left( \omega \right),
\end{equation}
where $*$ denotes the convolution operator. Since we assume a time-periodic excitation, all quantities have discrete spectra. The HBFEM formulation is then obtained from Eq.~\eqref{eq:MQS_NL_FD} by truncating the convolution after a given maximum harmonic order $m \in \mathbb{N}$ and subsequently discretizing according to the FE method. This leads to 
\begin{equation}\label{eq:HBFEM}
	\sum_{k = \max\{-m,n-m\} }^{\min\{m, n+m\}} \mathbf{K}_{\nu_k} \left( \bow{\underline{\mathbf{a}}}\right)\bow{\underline{\mathbf{a}}}_{n-k} + \jmath n\wf\mathbf{M}_{\sigma}\bow{\underline{\mathbf{a}}}_{n} = \secondbow{\bow{\underline{\mathbf{j}}}}_{\mathrm{s},n}
\end{equation}
for $ n \in \mathbb{Z} \, \cap \, [-m,m]$, where $\wf$ is the fundamental angular frequency, $\bow{\underline{\mathbf{a}}}_{n}$ the vector gathering the \mbox{d.o.f.} of the $n$-th harmonic of the magnetic vector potential, $\secondbow{\bow{\underline{\mathbf{j}}}}_{\mathrm{s},n}$ the discretized $n$-th harmonic of the source current density, $\mathbf{M}_{\sigma}$ the mass matrix, and  $\mathbf{K}_{\nu_k}$ denotes the stiffness matrix computed with the $k$-th harmonic of the reluctivity~\cite{roppert_2019aa}. The notation $\mathbf{K}_{\nu_k}\left(\bow{\underline{\mathbf{a}}}\right)$
indicates that each harmonic of the reluctivity depends on all harmonics of the magnetic vector potential. Considering that $\bow{\underline{\mathbf{a}}}_{-n} = \bow{\underline{\mathbf{a}}}_{n}^{*}$, we only have to solve for $\left( \bow{\underline{\mathbf{a}}}_{n} \right)_{n \in \mathbb{N}_0}$.
\subsection{Homogenization Technique}\label{sec:sim_method_hom}
\begin{figure}
	\centering
	\includegraphics[width=0.355\textwidth]{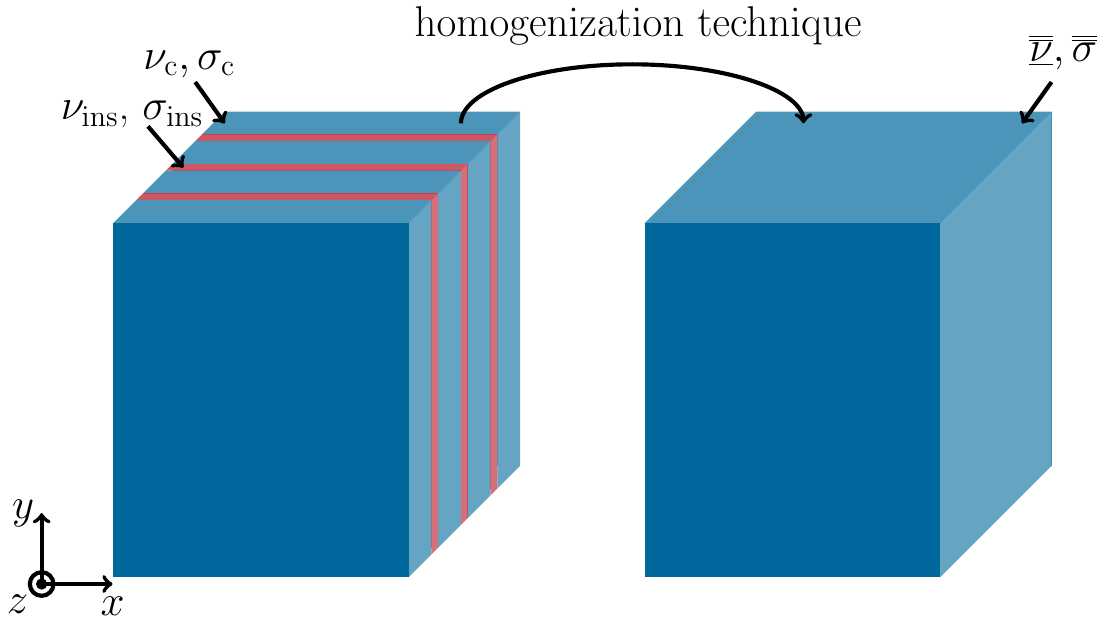}
	\caption{Left: Lamination stack with insulation in red and conducting laminates in blue. Right: Homogenized model.}
	\label{fig:homogenization}
\end{figure}

In the magnets' laminated yokes, the reluctivity and the conductivity depend on the spatial coordinate $\vec{r}$, since they are different for the conducting laminates and their insulating coating. Choosing the coordinate system as shown in Fig.~\ref{fig:homogenization}, the homogenization proposed in~\cite{dular_2003aa} transforms the laminated yokes into bulk models by replacing $\sigma(\vec{r})$ and $\nu(\vec{r})$ with spatially constant material tensors
\begin{align}
	\sigmatensor &= \sigma_{\mathrm{c}} \begin{bmatrix}
		1 & 0 & 0 \\ 0 & 1 & 0 \\ 0 & 0 & 0
	\end{bmatrix}, \label{eq:tensor_conductivity} \\
	\begin{split}
		\underline{\nutensor} &=  \frac{\sigma_{\mathrm{c}} d \delta \omega \left(1+\jmath\right)}{8} \frac{\sinh \left( \left(1 + \jmath \right) \frac{d}{\delta}\right)}{\sinh^2\left( \left(1 + \jmath \right) \frac{d}{2\delta}\right)} 
		\begin{bmatrix}
			1 & 0 & 0 \\ 0 & 1 & 0 \\ 0 & 0 & 0 
		\end{bmatrix} \\  & +  \nu_{\mathrm{c}} \begin{bmatrix}
			0 & 0 & 0 \\ 0 & 0 & 0 \\ 0 & 0 & 1
		\end{bmatrix},\label{eq:tensor_reluctivity}
	\end{split}
\end{align}
where $\sigma_{\mathrm{c}}$ denotes the conductivity of the laminates, $\nu_{\mathrm{c}}$ their reluctivity, $d$ their thickness, and $ \delta = \sqrt{\frac{2 \nu_{\mathrm{c}}}{\sigma_{\mathrm{c}}\omega}}$ is the skin depth.

\subsection{Homogenized Harmonic Balance Finite Element Method}
The HomHBFEM is obtained from Eq.~\eqref{eq:HBFEM} by introducing the homogenized material tensors. We write
\begin{align}\label{eq:homhbfem}
	\mathbf{K}_{\nutensor_0 \left(n\wf \right)}\left(\bow{\underline{\mathbf{a}}}\right)\bow{\underline{\mathbf{a}}}_{n} &+ \sum_{\substack{k= \max\{-m,n-m\}, \\ k \neq 0}}^{\min\{m, n+m\}} \mathbf{K}_{\nu_k} \left( \bow{\underline{\mathbf{a}}}\right)\bow{\underline{\mathbf{a}}}_{n-k} \notag \\ &+ \jmath n\wf\mathbf{M}_{\sigmatensor}\bow{\underline{\mathbf{a}}}_{n} = \secondbow{\bow{\underline{\mathbf{j}}}}_{\mathrm{s},n},
\end{align}
where $\mathbf{M}_{\sigmatensor}$ is the mass matrix constructed with the conductivity tensor given in Eq.~\eqref{eq:tensor_conductivity}, and $\mathbf{K}_{\nutensor_0 \left(n\wf \right)}$ is the stiffness matrix constructed with the reluctivity tensor given in Eq.~\eqref{eq:tensor_reluctivity} evaluated at $\omega = n \wf$ and $\nu = \nu_0$ for each element, where $\nu_0$ is the DC component of the reluctivity. 

The system in Eq.~\eqref{eq:homhbfem} is linearized via successive substitution. To avoid solving multiharmonic systems of equations increasing in size according to the number of considered harmonics, we adopt a block Jacobi iteration~\cite{saad_2003aa}, leading to 
\begin{align}\label{eq:homhbfem_iteration}
	& \left( \mathbf{K}_{\nutensor_0 \left(n\wf \right)}(\bow{\underline{\mathbf{a}}}^{i}) + \jmath n\wf\mathbf{M}_{\sigmatensor}   \right)    \bow{\underline{\mathbf{a}}}^{i + 1}_{n} \notag\\ &= \secondbow{\bow{\underline{\mathbf{j}}}}_{\mathrm{s},n} - \sum_{\substack{k= \max\{-m,n-m\}, \\ k \neq 0}}^{\min\{m, n+m\}} \mathbf{K}_{\nu_k} ( \bow{\underline{\mathbf{a}}}^{i})\bow{\underline{\mathbf{a}}}^{i}_{n-k}, 
\end{align}
where the superscript $i = 0,1,2,...$ indicates the iteration number. In this way, the method is easily parallelized, i.e., the equations for the different harmonics can be solved in parallel in each iteration~\cite{yamada_1991aa}.

To compute the stiffness matrices in each iteration, we first need to compute the discretized magnetic flux densities $\secondbow{\bow{\underline{\mathbf{b}}}}_{n}^{\raisebox{-5pt}{$\scriptstyle i$}}$ as the discrete curl of the $\bow{\underline{\mathbf{a}}}_{n}^{i}$. These complex-valued magnetic flux densities are transformed into the time domain. The time signal for the magnitude of the magnetic flux density is then inserted into the nonlinear $B-H$ curve to obtain the time signal for the magnitude of the magnetic field strength. By forming the quotient of these two time signals, we obtain the reluctivity~$\nu^{i}(t)$. Finally, computing the fast Fourier transform (FFT) of $\nu^{i}(t)$ gives us the Fourier series coefficients of the reluctivity, which allows us to set up $\mathbf{K}_{\nutensor_0 \left(n\wf \right)}(\bow{\underline{\mathbf{a}}}^{i})$ and $\mathbf{K}_{\underline{\nu}_k}(\bow{\underline{\mathbf{a}}}^{i})$ and to assemble the system of equations in Eq.~\eqref{eq:homhbfem_iteration}. 
\begin{figure}
	\centering
	\begin{tikzpicture}[node distance=1.5cm]
		\node (start) [start] {Initialize $\ahatc{n}^{0}, \secondbow{\bow{\underline{\mathbf{b}}}}_{n}^{\raisebox{-5pt}{$\scriptstyle 0$}}$};
		\node (pro1) [process_blue, below of=start, align=center] {Transform  $\secondbow{\bow{\underline{\mathbf{b}}}}_{n}^{\raisebox{-5pt}{$\scriptstyle i$}}$ into\\ time domain};
		\node (pro2) [process_blue, below of=pro1, align=center] {Compute $\nu^{i}\left(t\right)$ with\\nonlinear $B$-$H$ curve};
		
		\node (pro3) [process_blue, right=1.2cm of pro2, align=center] {FFT of $\nu^{i}\left( t\right)$};
		\node (pro4) [process_blue, above= 0.4cm of pro3, align=center] {Compute reluctivity \\ tensors $\nutensor_{0}^{i}\left(n\wf\right)$};
		\node (pro5) [process_red, above=0.4cm of pro4, align=center] {Assemble system Eq.~\eqref{eq:homhbfem_iteration} \\  \& solve  $\rightarrow \ahatc{n}^{i+1}, \secondbow{\bow{\underline{\mathbf{b}}}}_{n}^{\raisebox{-5pt}{$\scriptstyle i + 1$}}$};
		\node (dec1) [decision, above=0.4cm of pro5, node distance=1.5cm] {Convergence?};
		\node (stop) [stop, left=1.2cm of dec1, node distance=1.5cm] {Stop};
		
		\draw [arrow] (start) -- (pro1);
		\draw [arrow] (pro1) -- (pro2);
		\draw [arrow] (pro2) -- (pro3);
		\draw [arrow] (pro3) -- (pro4);
		\draw [arrow] (pro4) -- (pro5);
		\draw [arrow] (pro5) -- (dec1);
		\draw [arrow] (dec1) -- (stop);
		\node[anchor=west] at (2.0,1.85) {yes};
		\draw [arrow, rounded corners] (2.7,1.63) -- (2.7,-1.5) -- (1.75,-1.5);
		\node[rotate=90, anchor=west] at (2.55,0.8) {no};
		\node[anchor=west] at (1.9, -1.7) {$\scriptstyle i = i + 1$};
	\end{tikzpicture}
	\caption{Flow chart of the HomHBFEM. The index $n$ indicates the harmonic order and superscript $i$ the iteration number. Blue boxes are implemented in Python, red boxes in GetDP.}\label{fig:flowchart}
\end{figure}
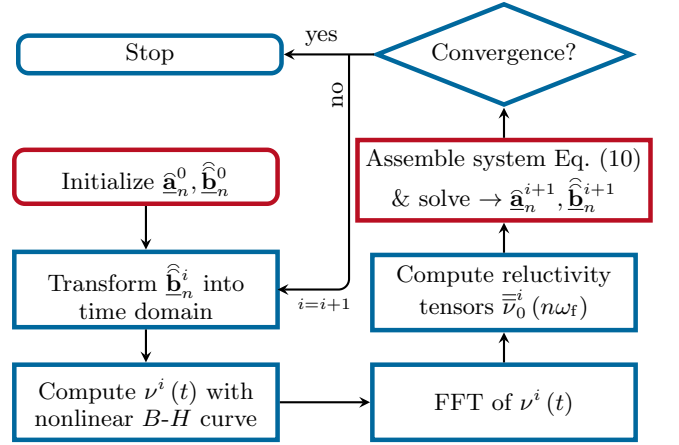

The iteration is stopped upon fulfillment of an energy-based convergence criterion. The described iterative procedure is illustrated in Fig.~\ref{fig:flowchart}. It has been implemented in the open-source FE software GetDP~\cite{dular_1999aa} combined with additional Python code.

\section{Measurement and Simulation Results}\label{sec:meas_vs_sim}
This section presents the experimental validation of the HomHBFEM. First, the simulation results are compared to the Hall sensor measurements, then to the search coil measurements of the prototype FC magnet.
\subsection{DC Investigation via Hall Sensor}
\begin{figure}
    \centering
    \includegraphics[width=0.5\linewidth]{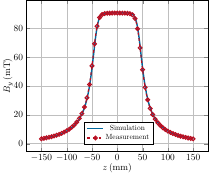}
    \caption{Comparison of simulated and measured values of the DC magnetic flux density along the longitudinal axis of the FC magnet.}
    \label{fig:meas_vs_sim_1}
\end{figure}

Before we begin analyzing the AC results, we first check the DC results for the longitudinal field profile, i.e., the magnitude of the magnetic flux density along the axis, to confirm the correctness of the simulation model and the proper configuration of the experimental setup. Figure~\ref{fig:meas_vs_sim_1} shows the longitudinal field profile at the maximum DC current of $I_{\mathrm{DC}} = \SI{15}{\ampere}$. The simulation has been performed using the measured nonlinear $B$--$H$ curve shown in Fig.~\ref{fig:fc_magnet_2}, both in CST Studio Suite~\cite{cst} and GetDP~\cite{dular_1999aa}, with practically identical results. The measurement has been performed with a different Hall sensor than the one for the AC measurements described in Sec.~\ref{sec:meas_setup}, with a larger range for the magnetic flux density. Measurement and simulation show excellent agreement. 
\subsection{AC Investigation via Hall Sensor}\label{sec:sim_vs_meas_hall_ac}

\subsubsection{Longitudinal Field Profiles}
\begin{figure}
  \centering

  \begin{minipage}[t]{0.48\columnwidth}
    \centering
    \includegraphics[width=\linewidth]{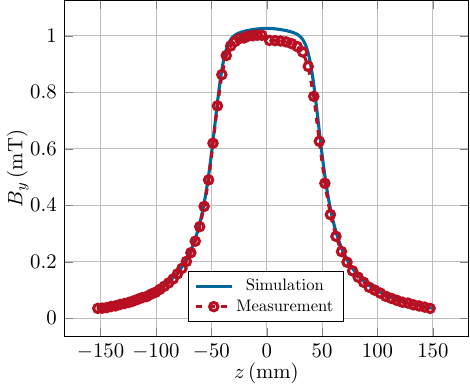}
    \par\smallskip
    \hspace{8mm}(a)
    \label{fig:meas_vs_sim_2_1}
  \end{minipage}\hfill
  \begin{minipage}[t]{0.48\columnwidth}
    \centering
    \includegraphics[width=\linewidth]{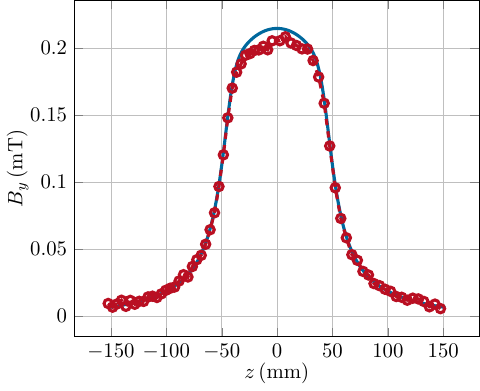}
    \par\smallskip
    \hspace{8mm}(b)
    \label{fig:meas_vs_sim_2_2}
  \end{minipage}

  \vspace{2mm}

  \begin{minipage}[t]{0.48\columnwidth}
    \centering
    \includegraphics[width=\linewidth]{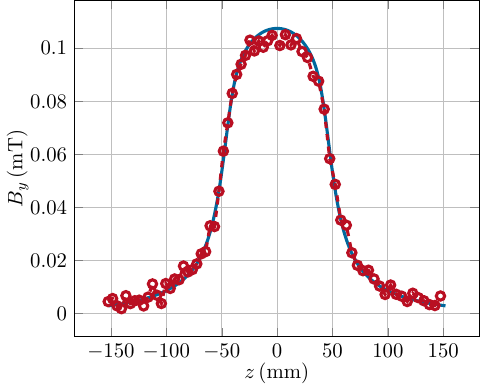}
    \par\smallskip
    \hspace{8mm}(c)
    \label{fig:meas_vs_sim_2_3}
  \end{minipage}\hfill
  \begin{minipage}[t]{0.48\columnwidth}
    \centering
    \includegraphics[width=\linewidth]{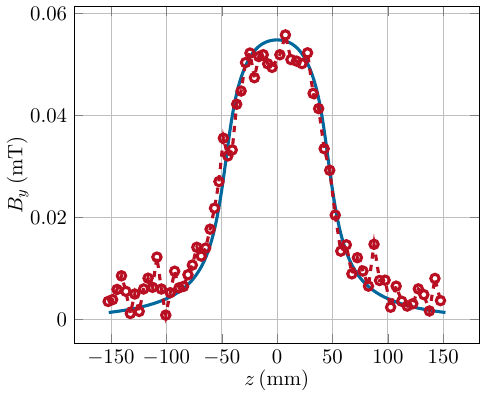}
    \par\smallskip
    \hspace{8mm}(d)
    \label{fig:meas_vs_sim_2_4}
  \end{minipage}

  \caption{Comparison of simulated and measured values for the magnitude of the magnetic flux density along the longitudinal axis of the FC magnet:
  (a) $f_{\mathrm{f}}=\SI{100}{\hertz}$,
  (b) $f_{\mathrm{f}}=\SI{500}{\hertz}$,
  (c) $f_{\mathrm{f}}=\SI{1}{\kilo\hertz}$,
  and (d) $f_{\mathrm{f}}=\SI{2}{\kilo\hertz}$.}
  \label{fig:meas_vs_sim_2}
\end{figure}

\begin{figure}
  \centering

  \begin{minipage}[t]{0.48\columnwidth}
    \centering
    \includegraphics[width=\linewidth]{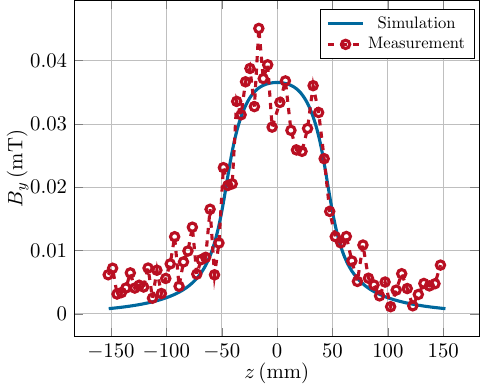}
    \par\smallskip
    \hspace{8mm}(a)
    \label{fig:meas_vs_sim_3_1}
  \end{minipage}\hfill
  \begin{minipage}[t]{0.48\columnwidth}
    \centering
    \includegraphics[width=\linewidth]{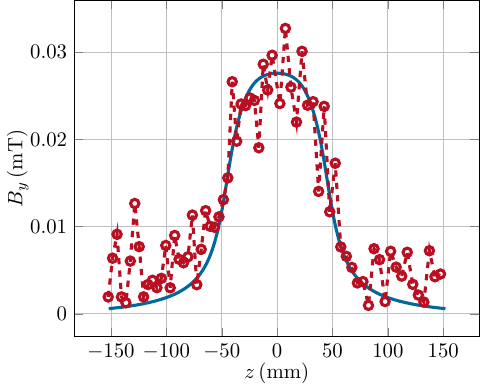}
    \par\smallskip
    \hspace{8mm}(b)
    \label{fig:meas_vs_sim_3_2}
  \end{minipage}

  \caption{Comparison of simulated and measured values for the magnitude of the magnetic flux density along the longitudinal axis of the FC magnet:
  (a) $f_{\mathrm{f}}=\SI{3}{\kilo\hertz}$ and
  (b) $f_{\mathrm{f}}=\SI{4}{\kilo\hertz}$.}
  \label{fig:meas_vs_sim_3}
\end{figure}

The comparison of the HomHBFEM simulation results and the measurements of the AC longitudinal field profiles is shown in Fig.~\ref{fig:meas_vs_sim_2} for excitation frequencies between \mbox{$\ff = \SI{100}{\hertz}$} and \mbox{$\ff = \SI{2}{\kilo\hertz}$}. For the HomHBFEM results, we always show the first harmonic of the magnetic flux density, higher order harmonics are negligible. We observe good agreement between measurement and simulation. At frequencies around $\SI{3}{\kilo\hertz}$ - $\SI{4}{\kilo\hertz}$, the position-resolved measurements become too noisy to properly compare to the simulation results, see Fig.~\ref{fig:meas_vs_sim_3}. 

\begin{figure}
    \centering
    \includegraphics[width=0.7\linewidth]{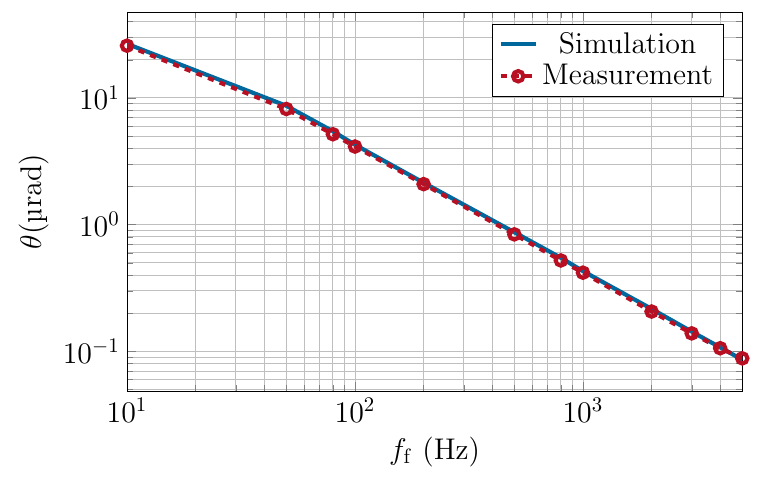}
    \caption{Comparison of AC deflection angles computed from measurements and simulations as a function of frequency.}
    \label{fig:meas_vs_sim_deflection angle}
\end{figure}

Nevertheless, the results in this higher frequency range are still promising: The overall trend of the measurements remains consistent with the simulation and, most importantly, the integrated values, which are of particular interest since they define the deflection angle $\theta$, still match well, even up to the maximum measurement frequency of $\SI{5}{\kilo\hertz}$. This is shown in Fig.~\ref{fig:meas_vs_sim_deflection angle}, which compares the deflection angle computed from the measured and simulated field profiles via
\begin{equation}
    \theta(\ff) = \frac{c}{E}\int_{\SI{-43}{\milli\meter}}^{\SI{43}{\milli\meter}} B_y(\ff,z) \mathrm{d}z
\end{equation}
where $B_y$ denotes the magnitude of the vertical component of the magnetic flux density (the main field component in this case), $c$ is the speed of light, and $E$ the total electron energy in $\si{\electronvolt}$~\cite{wiedemann_2015aa}, which in the case of PETRA IV is $\SI{6}{\giga\electronvolt}$. For the sake of the comparison between measurement and simulation, the integration limits are chosen at the start and end point of the magnet yoke, i.e., at $z = \pm \SI{43}{\milli\meter}$, since the measurement results above $\SI{2}{\kilo\hertz}$ outside of the yoke are dominated by noise, see Fig.~\ref{fig:meas_vs_sim_3}. The average relative deviation of the deflection angle computed from the HomHBFEM simulation compared to the deflection angle computed from the measurements between $\SI{10}{\hertz}$ and $\SI{5}{\kilo\hertz}$ is $\SI{3.6}{\percent}$ and the maximum relative deviation is $\SI{6.0}{\percent}$. 

\subsubsection{Importance of the Initial Permeability}
It was realized during the measurement campaign that, if special care is taken and if the excitation frequency is sufficiently high, one can also achieve good agreement in the field profiles using the original homogenization technique, as given in Sec.~\ref{sec:sim_method_hom}.
Unlike the HomHBFEM, the original homogenization does not consider the nonlinear $B$--$H$ curve but requires to set a constant value for the permeability. This value must be chosen carefully. As we will explain in the following, the initial permeability of the laminates should be used. 

The strength of the AC corrections that must be delivered by the FC magnet declines with increasing frequency. As a result, the current and magnetic field amplitudes at high frequencies are low. Therefore, the low-field region of the nonlinear $B$--$H$ curve of the laminates is of particular interest to us. 

\begin{figure}
  \centering
  \begin{minipage}[t]{0.48\columnwidth}
    \centering
    \includegraphics[width=\linewidth]{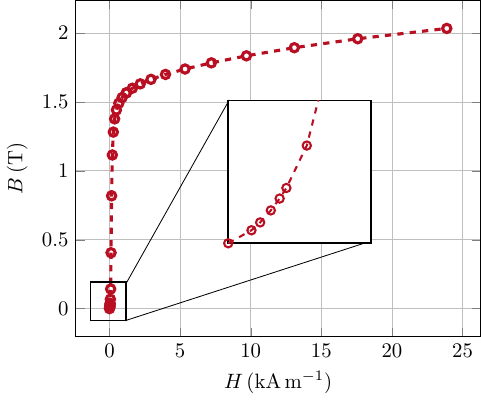}
    \caption{$B$--$H$ curve with close-up of Rayleigh region.}
    \label{fig:meas_vs_sim_rayleigh_1}
  \end{minipage}\hfill
  \begin{minipage}[t]{0.5\columnwidth}
    \centering
    \includegraphics[width=\linewidth]{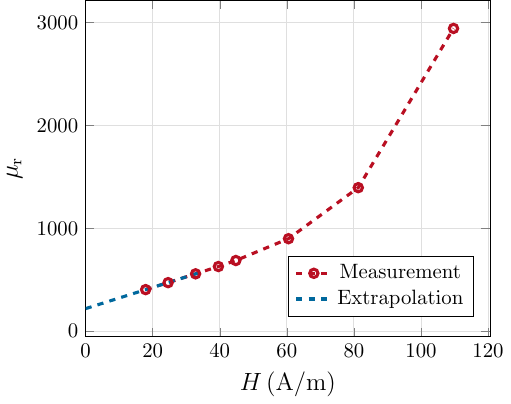}
    \caption{Relative permeability with extrapolation.}
    \label{fig:meas_vs_sim_rayleigh_2}
  \end{minipage}
\end{figure}

Ferromagnetic materials contain defects and impurities. As a consequence, the relation between the magnetic flux density $B$ and the magnetic field strength $H$ at low field strengths is not linear, but quadratic~\cite{kronmuller_2003aa}. The corresponding region in the $B$--$H$ curve is commonly referred to as the Rayleigh region. In this region, which typically extends from $H = \SI{0}{\ampere\per\meter}$ to roughly $\SI{80}{\ampere\per\meter}$~\cite{cullity_2009aa}, the \mbox{$B$--$H$} curve can be described by
\begin{equation}
    B(H) = \mu_{\mathrm{init}}H + \eta H^2,\label{eq:rayleigh_law_intro}
\end{equation}
where $\mu_{\mathrm{init}}$ and $\eta$ are constants and $\mu_{\mathrm{init}}$ is the initial permeability, i.e., the permeability at $H = 0$. 
A close-up of the Rayleigh region of the $B$--$H$ curve of \mbox{powercore$\textsuperscript{\textregistered}$ 1400-100AP} is shown in Fig.~\ref{fig:meas_vs_sim_rayleigh_1}.

The initial permeability is typically not directly measured, since measurements at such low field strengths are difficult. Instead, it is determined via the following standard procedure~\cite{bozorth_1993aa}. We plot the measurements for the permeability over $H$ and then extrapolate down to $H=0$ by fitting a straight line to the last few measurement points. This is illustrated in Fig.~\ref{fig:meas_vs_sim_rayleigh_2} for \mbox{powercore$\textsuperscript{\textregistered}$ 1400-100AP}. In this way, we obtain an estimated initial relative permeability of $\mu_{\mathrm{r,init}} \approx 220$. Note that the described procedure may overestimate $\mu_{\mathrm{r,init}}$ for certain materials where the \mbox{$\mu$--$H$} curve bends downward for $H \rightarrow 0$ and thus deviates from the linear extrapolation~\cite{sixtus_1943aa,cullity_2009aa}.

\begin{figure}
  \centering

  \begin{minipage}[t]{0.48\columnwidth}
    \centering
    \includegraphics[width=\linewidth]{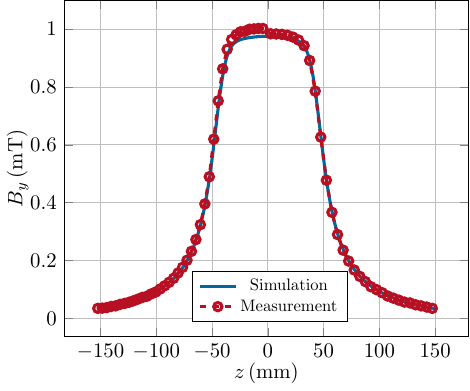}
    \par\smallskip
    \hspace{8mm}(a)
    \label{fig:meas_vs_sim_5_1}
  \end{minipage}\hfill
  \begin{minipage}[t]{0.48\columnwidth}
    \centering
    \includegraphics[width=\linewidth]{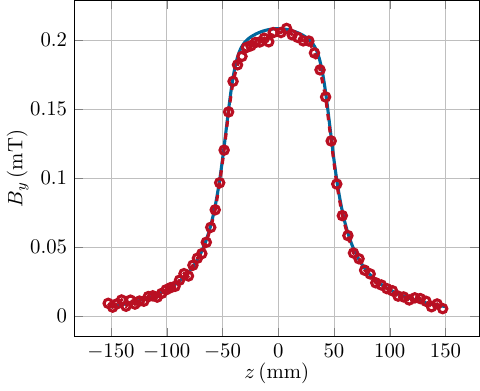}
    \par\smallskip
    \hspace{8mm}(b)
    \label{fig:meas_vs_sim_5_2}
  \end{minipage}

  \vspace{2mm}

  \begin{minipage}[t]{0.48\columnwidth}
    \centering
    \includegraphics[width=\linewidth]{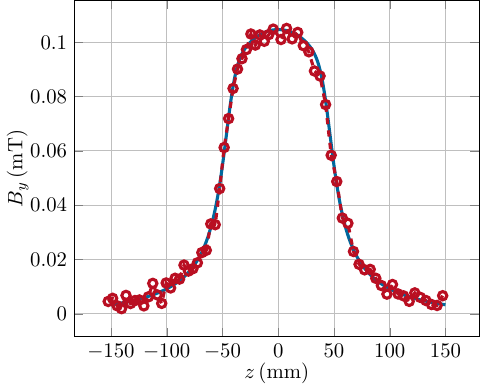}
    \par\smallskip
    \hspace{8mm}(c)
    \label{fig:meas_vs_sim_5_3}
  \end{minipage}\hfill
  \begin{minipage}[t]{0.48\columnwidth}
    \centering
    \includegraphics[width=\linewidth]{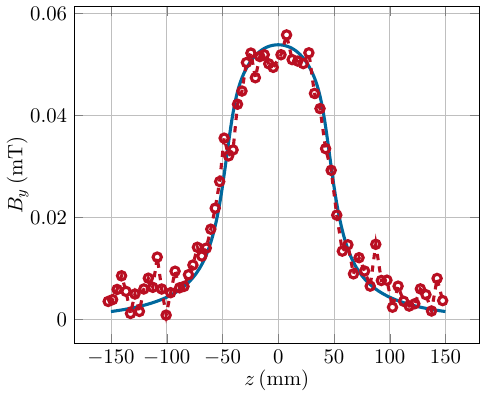}
    \par\smallskip
    \hspace{8mm}(d)
    \label{fig:meas_vs_sim_5_4}
  \end{minipage}

  \caption{Comparison of simulated and measured values for the magnitude of the magnetic flux density along the longitudinal axis of the FC magnet. The simulation uses the initial permeability. 
  (a) $f_{\mathrm{f}} = \SI{100}{\hertz}$, 
  (b) $f_{\mathrm{f}} = \SI{500}{\hertz}$, 
  (c) $f_{\mathrm{f}} = \SI{1}{\kilo\hertz}$, and 
  (d) $f_{\mathrm{f}} = \SI{2}{\kilo\hertz}$.}
  \label{fig:meas_vs_sim_5}
\end{figure}

\begin{figure}
  \centering

  \begin{minipage}[t]{0.48\columnwidth}
    \centering
    \includegraphics[width=\linewidth]{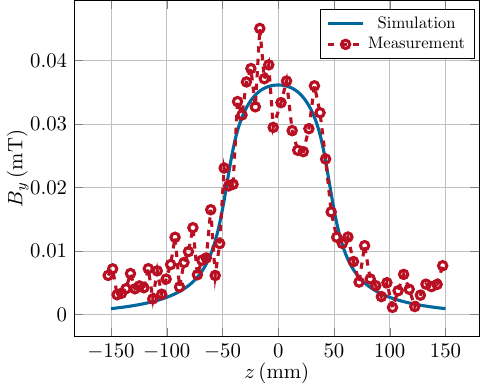}
    \par\smallskip
    \hspace{8mm}(a)
    \label{fig:meas_vs_sim_6_1}
  \end{minipage}\hfill
  \begin{minipage}[t]{0.48\columnwidth}
    \centering
    \includegraphics[width=\linewidth]{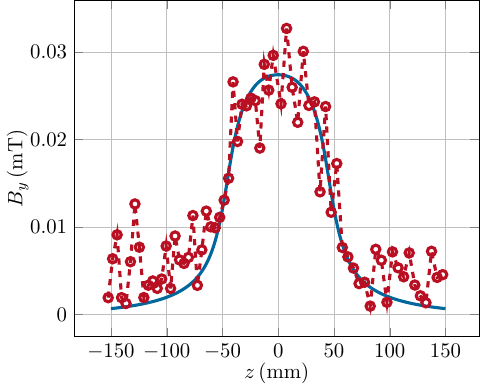}
    \par\smallskip
    \hspace{8mm}(b)
    \label{fig:meas_vs_sim_6_2}
  \end{minipage}

  \caption{Comparison of simulated and measured values for the magnitude of the magnetic flux density along the longitudinal axis of the FC magnet. The simulation uses the initial permeability. 
  (a) $f_{\mathrm{f}} = \SI{3}{\kilo\hertz}$ and 
  (b) $f_{\mathrm{f}} = \SI{4}{\kilo\hertz}$.}
  \label{fig:meas_vs_sim_6}
\end{figure}

The simulated longitudinal field profiles using this value for the initial permeability are compared with the measured results in Fig.~\ref{fig:meas_vs_sim_5} for frequencies between \mbox{$\ff = \SI{100}{\hertz}$} and \mbox{$\ff = \SI{2}{\kilo\hertz}$}. For completeness, Fig.~\ref{fig:meas_vs_sim_6} again shows the comparison for $\ff = \SI{3}{\kilo\hertz}$ and $\SI{4}{\kilo\hertz}$, where the measurements become increasingly noisy. Overall, the simulation results using the original homogenization with the initial permeability are very similar to those with the HomHBFEM using the \mbox{$B$--$H$ curve}. Both show good a agreement with the measurements for the investigated frequency range. At the lower frequencies, however, where the current amplitude is higher, significant differences become visible. This will be shown in the next subsection.

\subsubsection{Quasi-DC Regime}
For low frequencies, i.e., below $\SI{100}{\hertz}$, the use of the original homogenization with the initial permeability is not fully justified since the currents and resulting field strengths in the prototype FC magnet yoke are not sufficiently small. Hence, the permeability of the yoke is in this case significantly higher than the initial permeability and as a result, the simulation using the initial permeability underestimates the magnetic flux density in the aperture.

\begin{figure}
  \centering

  \begin{minipage}[t]{0.48\columnwidth}
    \centering
    \includegraphics[width=\linewidth]{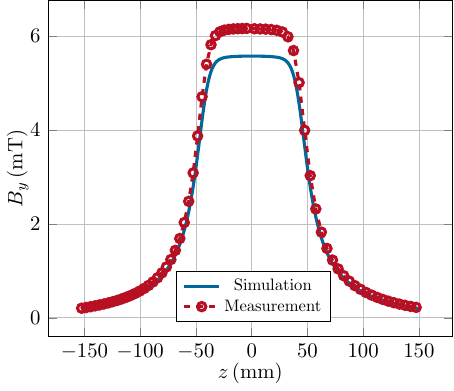}
    \par\smallskip
    \hspace{8mm}(a)
    \label{fig:meas_vs_sim_quasi_dc_1}
  \end{minipage}\hfill
  \begin{minipage}[t]{0.48\columnwidth}
    \centering
    \includegraphics[width=\linewidth]{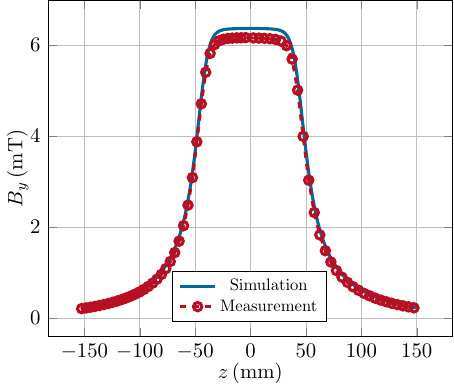}
    \par\smallskip
    \hspace{8mm}(b)
    \label{fig:meas_vs_sim_quasi_dc_2}
  \end{minipage}

  \caption{Comparison of simulated and measured values for the magnitude of the magnetic flux density along the longitudinal axis of the FC magnet at $f_{\mathrm{f}} = \SI{10}{\hertz}$. 
  (a) Simulation with initial permeability and 
  (b) simulation with HomHBFEM.}
  \label{fig:meas_vs_sim_quasi_dc}
\end{figure}

Figure \ref{fig:meas_vs_sim_quasi_dc}(a) illustrates this for the lowest measurement frequency, $\ff = \SI{10}{\hertz}$, where we recorded the largest differences between the simulation approach with the initial permeability and the measurements. In this case, the maximum relative deviation between simulation and measurement in the center ($z = 0$) of the prototype FC magnet, is $\SI{10}{\percent}$.

Figure~\ref{fig:meas_vs_sim_quasi_dc}(b) shows that this problem can be largely alleviated by simulating with the HomHBFEM approach at low frequencies, using the nonlinear $B$--$H$ curve instead of a fixed permeability value. As expected based on the explanation above, the simulated magnetic flux densities in the aperture are now significantly higher than in the approach with the initial permeability. In fact, they are now slightly larger than the measured ones, but the relative error in the center of the magnet is significantly reduced to roughly $\SI{3}{\percent}$. The remaining discrepancy is most likely due to one or a combination of the following factors: inaccuracies of the \mbox{$B$--$H$} curve used in the simulation, particularly at low fields, a slight tilt of the Hall probe, and/or some additional damping by an element of the measurement chain.

\subsubsection{Integrated Transfer Function} 
For the design of the feedback control, we are primarily interested in the damping of the integrated magnetic flux density with increasing frequency relative to the supplied current as well as the phase shift between the magnetic flux density and the supplied current. To that end, we investigate the integrated transfer function (ITF) defined as
\begin{equation}\label{eq:itf}
    \underline{G}_{\mathrm{int}}(\jmath \wf) = \left. \frac{\int \underline{B}(z,\wf) \d z}{\int B_{\mathrm{DC}}(z)\d z} \middle/ \frac{\underline{I}(\wf) }{I_{\mathrm{DC}}} \right. ,
\end{equation}
where $\underline{I}(\wf)$ and $\int \underline{B}(z,\wf) \d z$ are the current amplitude and the integrated dipole field at a given excitation frequency and $I_{\mathrm{DC}}$ and $\int B_{\mathrm{DC}}(z) \d z$ are the respective \mbox{DC quantities}.

\begin{figure}
    \centering
    \includegraphics[width=\linewidth]{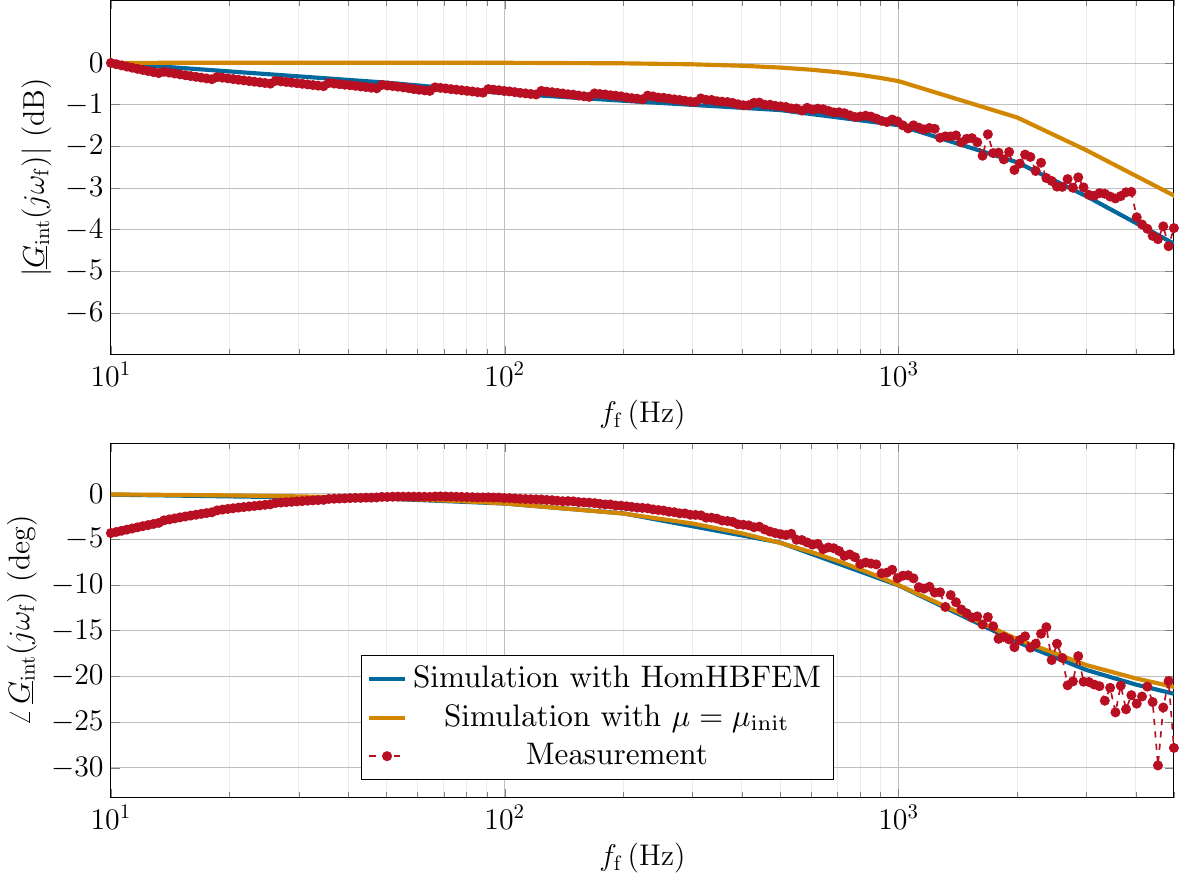}
    \caption{Magnitude and phase of the ITF. Hall sensor measurement compared to the two simulation approaches.}
    \label{fig:meas_vs_sim_8}
\end{figure}

The ITFs computed from the measured and simulated magnitude and phase of the magnetic flux densities along the axis are shown in Fig.~\ref{fig:meas_vs_sim_8}. Note that we normalized to the results at the lowest measurement frequency \mbox{$\ff = \SI{10}{\hertz}$}, i.e., the DC quantities in Eq.~\eqref{eq:itf} were replaced by the corresponding quantities at $\SI{10}{\hertz}$.

Just like for the computation of the deflection angles, we again choose to integrate only from $z = - \SI{43}{\milli\meter}$ to $z = \SI{43}{\milli\meter}$, i.e., to only consider the magnetic flux density within the boundaries of the yoke. This is done to limit the effect of the measurement noise at the higher frequencies and thus allow for a better comparison between simulation and measurement.

Looking at the magnitude of the ITFs in Fig.~\ref{fig:meas_vs_sim_8}, we immediately notice that the HomHBFEM matches the measured ITF much better than the simulation with the original homogenization with the initial permeability. This is due to the fact that the magnitude of the measured ITF decreases already in the quasi-DC regime, i.e., for $\ff \leq \SI{100}{\hertz}$. This damping effect is not directly caused by the eddy currents in the yoke. Instead, this is related to the nonlinearity of the \mbox{$B$--$H$} curve:  As the frequency is increased, the current in the measurement setup decreases due to the increasing impedance of the magnet. Consequently, the field strength in the yoke decreases and thus the permeability of the yoke declines, see Fig.~\ref{fig:fc_magnet_2}, which leads to a decrease in the magnetic flux density in the aperture.

Since the simulation with the original homogenization with the initial permeability uses a constant permeability, it cannot capture the described behavior at low frequencies. Consequently, the simulation result for the magnitude of the ITF with this approach deviates significantly from the measurement result. 

With the HomHBFEM, on the other hand, the nonlinear \mbox{$B$--$H$} curve is included in the simulation and thus the change of the permeability is considered. As a result, it can capture the described behavior at the lower frequencies, which leads to a very good agreement with the measurement result for the magnitude of the ITF.

The measured phase shows some unexpected behavior at frequencies below $\SI{30}{\hertz}$. At such low frequencies, the eddy-current effects should be negligible thanks to the
lamination of the yoke and hence the phase is expected to be zero. Both the HomHBFEM and the simulation with the initial permeability yield results that are in agreement with this expectation, whereas the measurements show a phase shift of roughly $\SI{-4.3}{\degree}$ at $\SI{10}{\hertz}$. The origin of this unexpected behavior is unclear and will be further studied. At frequencies above $\SI{30}{\hertz}$, however, the agreement between the measurement and both simulation approaches is good. The deviations between measurement and HomHBFEM simulation results for the ITF are quantified in Table~\ref{tab:meas_vs_sim_hall_sensor_itf}.

The time needed to simulate the magnet with the HomHBFEM depends on the level of the excitation current. With the currents used here, see Fig.~\ref{fig:hall_sensor_meas_current}, we need roughly 10 min per frequency point on a standard workstation. Consequently, computing the ITF over the full frequency range requires a few hours of simulation time. By comparison, the simulation with the original homogenization with constant permeability is much faster, on the order of a minute per frequency point, such that the ITF can be computed in less than half an hour. 

\setlength{\tabcolsep}{8pt}  
\begin{table}
\caption{Maximum and average deviation of the HomHBFEM simulation results for the ITF from the measurements with the Hall sensor.}\label{tab:meas_vs_sim_hall_sensor_itf}
\centering
\begin{tabular}{lcc}
    \toprule
    \toprule
     & Max. Deviation & Av. Deviation \\
    \midrule
    Mag. (dB)   & $0.69$ & $0.13$ \\
    Phase (deg) & $8.27$ & $1.12$    \\
    \bottomrule
    \bottomrule
\end{tabular}

\end{table}

\subsection{AC Investigation via Search Coils}\label{subsec:meas_vs_sim_search_coil}
\begin{figure}[t]
    \centering
    \includegraphics[width=\linewidth]{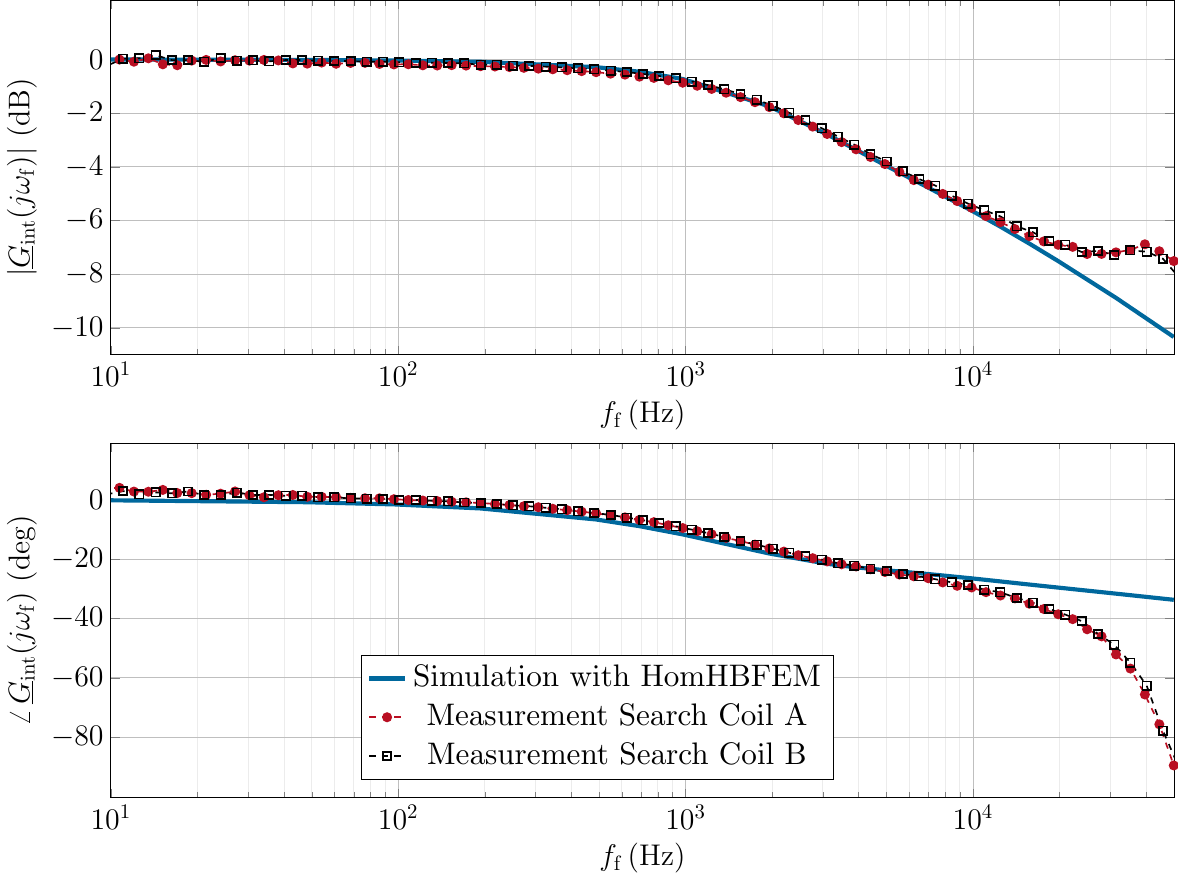}
    \caption{Magnitude and phase of the ITF from $\SI{10}{\hertz}$ to $\SI{50}{\kilo\hertz}$. Search coil measurements compared to HomHBFEM.}
    \label{fig:meas_vs_sim_9}
\end{figure}

Next, we compare the HomHBFEM simulation results to the measurements with the search coils. Since the search coils only give us the integrated magnetic flux densities and not the field profiles which we analyzed above, we show here directly the ITFs, see Fig.~\ref{fig:meas_vs_sim_9}. Note that in this case, the limits of the integrals in Eq.~\eqref{eq:itf} are set to $z = \pm \SI{150}{\milli\meter}$, since the search coils have a total length of $l_z = \SI{300}{\milli\meter}$. We observe a very good agreement between measurement and simulations for excitation frequencies up to roughly $\ff = \SI{10}{\kilo\hertz}$. At higher frequencies, the measurements are affected by a resonance, preventing a reliable comparison. 

\begin{figure}[H]
    \centering
    \includegraphics[width=\linewidth]{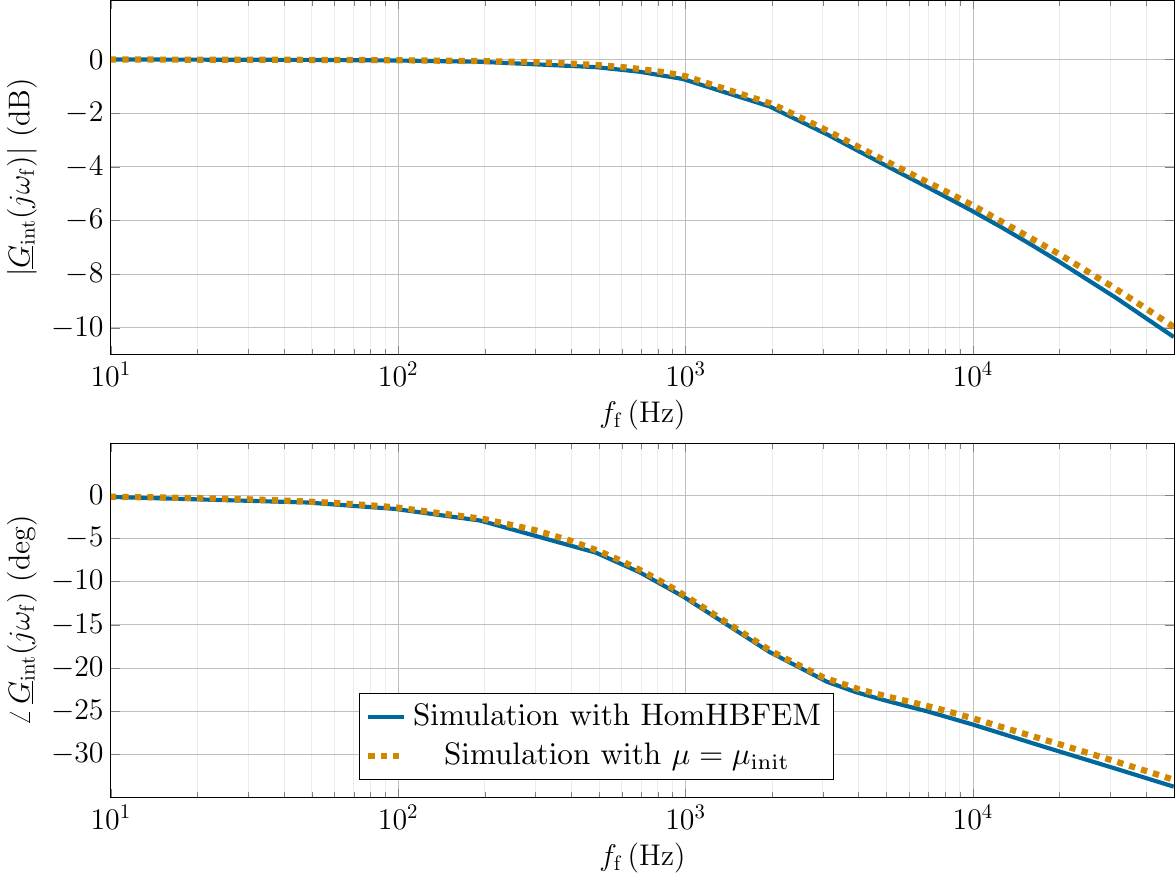}
    \caption{Magnitude and phase of the ITF from $\SI{10}{\hertz}$ to $\SI{50}{\kilo\hertz}$. HomHBFEM compared to the original homogenization technique with the initial permeability.}
    \label{fig:appendix_2}
\end{figure}

As explained in Sec.~\ref{sec:measurement_setup_search_coil}, for this second measurement with the search coils, the currents in the FC magnet coils are significantly lower than in the first measurement with the Hall sensor. As a result, nonlinear effects do not play a significant role. Hence, we do not see the same downward slope at the lower frequencies that we have seen in the magnitude of the ITF computed from the position-dependent measurements with the Hall sensor. Therefore, using the currents supplied to the magnet coils during the search coil measurement as excitation currents in the simulation, we obtain virtually identical results with the HomHBFEM and the simulation with the original homogenization using the initial permeability. This is shown in Fig.~\ref{fig:appendix_2}. 

\begin{figure}[H]
    \centering
    \includegraphics[width=\linewidth]{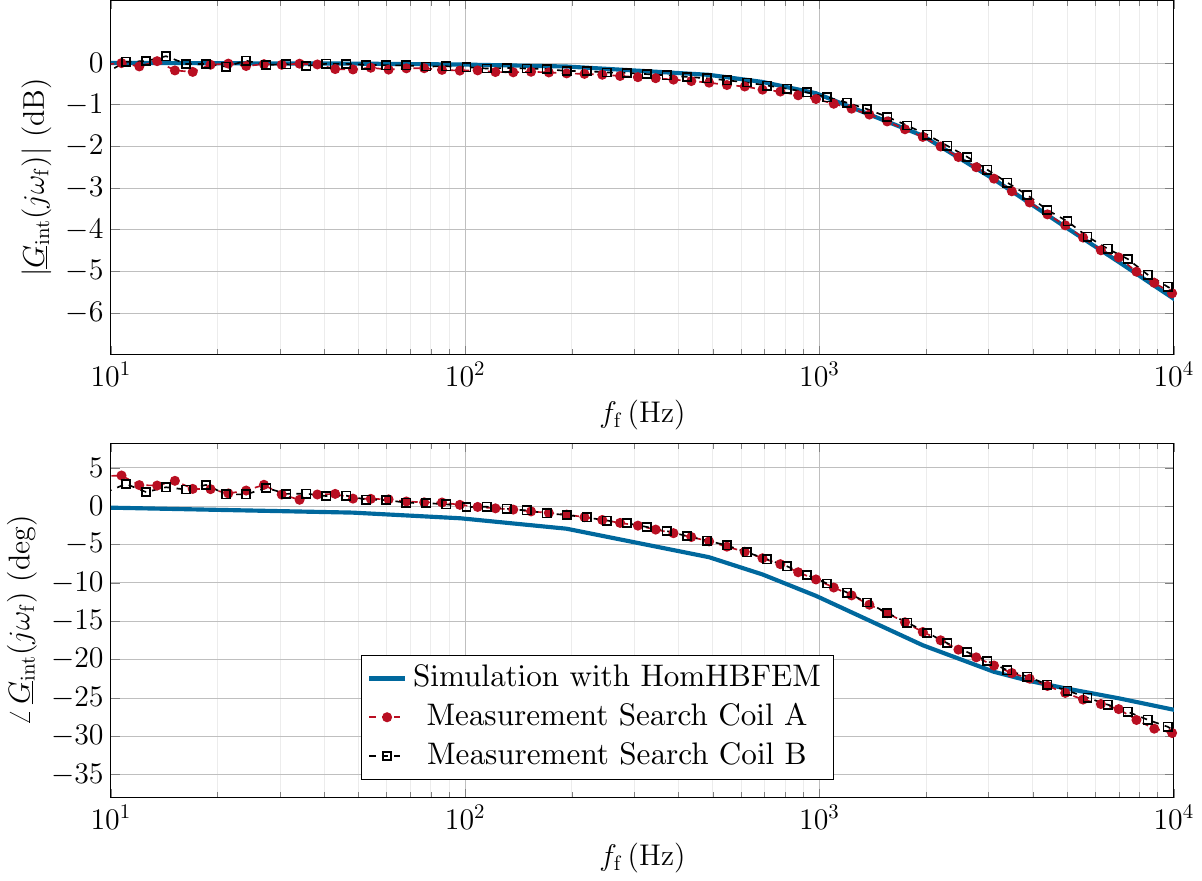}
    \caption{Magnitude and phase of the ITF from $\SI{10}{\hertz}$ to $\SI{10}{\kilo\hertz}$. Search coil measurements compared to HomHBFEM.}
    \label{fig:meas_vs_sim_10}
\end{figure}

It must be pointed out that while the agreement between simulation and measurement in the magnitude is excellent up until the resonance sets in, a closer look at the phase does reveal some deviations. For a more detailed analysis of these deviations, we focus on the frequency range from $\ff = \SI{10}{\hertz}$ to $\SI{10}{\kilo\hertz}$, i.e., we ignore the results at the frequencies that are significantly affected by the resonance. The corresponding close-up of the ITFs is shown in Fig.~\ref{fig:meas_vs_sim_10}. 
The maximum and averaged deviation of the HomHBFEM simulation results from the measurements with the two search coils are summarized in Table~\ref{tab:search_coil_1}. It should be highlighted here that for both search coils, the maximum deviation from the HomHBFEM simulation results in the phase occurs at very low frequencies (below $\SI{20}{\hertz}$). As discussed in Sec.~\ref{sec:sim_vs_meas_hall_ac}, the phase is expected to be zero (or slightly negative) at such low frequencies. The HomHBFEM simulation result for the phase is in agreement with this expectation, but the measurements show non-negligible positive values of up to $\SI{4.0}{\degree}$ for search coil A and $\SI{2.9}{\degree}$ for search coil B. These positive values for the phase are most likely due to a problem with the measurement setup and will be investigated further in the future. 

Despite the remaining issues with the measured phase in the quasi-DC regime, the overall very good agreement between simulated and measured ITFs over a broad frequency range from $\SI{10}{\hertz}$ to $\SI{10}{\kilo\hertz}$ further validates the HomHBFEM as a useful tool in the characterization of the dynamic behavior of the FC magnets. Furthermore, we conclude that the simulation approach with the original homogenization, using the initial permeability of the lamination material, can be a viable, easy-to-implement alternative for FC magnet simulations if the AC current amplitudes are sufficiently small. 
\setlength{\tabcolsep}{8pt}  
\begin{table}[H]
\caption{Maximum and average deviation of the HomHBFEM simulation results for the ITF from the measurements with the two search coils.}\label{tab:search_coil_1}
\centering
\begin{tabular}{lcccc}
\toprule
\toprule
 & \multicolumn{2}{c}{Max. Deviation} & \multicolumn{2}{c}{Av. Deviaton} \\
\cmidrule(lr){2-3} \cmidrule(lr){4-5}
 & Coil A & Coil B & Coil A & Coil B \\
\midrule
Mag.\ (dB)   & $0.21$ & $0.21$  & $0.10$  & $0.08$  \\
Phase (deg) & $4.20$ & $3.12$  & $1.94$ & $1.86$  \\
\bottomrule
\bottomrule
\end{tabular}

\end{table}

\section{Conclusion}\label{sec:conclusion}
We have presented a first experimental validation for the HomHBFEM, a dedicated simulation method which we have developed for efficient nonlinear eddy-current simulations of the FC magnets for PETRA IV. The HomHBFEM combines a frequency-domain-based homogenization technique with the HBFEM. In this way, the laminated yokes are replaced with bulk models that can be simulated with a relatively coarse FE mesh and computationally expensive time-stepping of the eddy-current field problem is avoided.

The experimental validation was performed by investigating the first prototype FC magnet for PETRA IV using two different measurement setups. In the first measurement setup, the magnetic flux density in the FC magnet aperture was measured with a Hall sensor that was moved along the axis. In the second measurement setup, the integrated magnetic flux density was measured with two different search coils.

In the measurement setup with the Hall sensor, we were able to show a good agreement between simulation and measurement in the ITF for excitation frequencies from $\SI{10}{\hertz}$ up to $\SI{5}{\kilo\hertz}$. In the measurement setup with the search coils, we were able to show a good agreement in the ITFs even up to $\SI{10}{\kilo\hertz}$.  Additionally, we have shown that if the amplitudes of the AC currents in the FC magnet's coils are sufficiently small, one may use the homogenization technique employed within the HomHBFEM in its original form, but with the constant permeability chosen as the initial permeability of the yoke laminates.

Further investigations will focus on alleviating the remaining issues with the phase measurements particularly at lower frequencies and exploring the behavior of the FC magnets at larger AC current amplitudes and under DC bias.  First investigations comparing simulation and measurements for the dynamic behavior of the magnet including a vacuum chamber have also been started and will be completed in the future.

\newpage
\bibliography{ref}

\end{document}